\documentclass[
  journal=bjps,
]{cup-journal}

\usepackage[frozencache,cachedir=.]{minted} \usepackage{booktabs,microtype,siunitx,tabularx}
\usetikzlibrary{trees}
\usepackage{parskip}
\usepackage{makecell}

\title{A Python Package for Sampling from Copulae: clayton}
\shorttitle{A Python Package for Sampling from Copulae: clayton}

\author{Alexis Boulin}
\affiliation{Université Côte d’Azur, CNRS, LJAD, France}
\alsoaffiliation{Inria, Lemon}

\begin{document}

\begin{abstract}
The package \textsf{clayton} is designed to be intuitive, user-friendly, and efficient. It offers a wide range of copula models, including Archimedean, Elliptical, and Extreme. The package is implemented in pure \textsf{Python}, making it easy to install and use. In addition, we provide detailed documentation and examples to help users get started quickly. We also conduct a performance comparison with existing \textsf{R} packages, demonstrating the efficiency of our implementation. The \textsf{clayton} package is a valuable tool for researchers and practitioners working with copulas in \textsf{Python}.
\end{abstract}

\section{Introduction\label{sec:introduction}}

    Modeling dependence relations between random variables is a topic of interest in probability theory and statistics. The most popular approach is based on the second moment of the underlying random variables, namely, the covariance. It is well known that only linear dependence can be captured by the covariance and it is only characteristic for a few models, e.g., the multivariate normal distribution or binary random variables. As a beneficial alternative to dependence, the concept of copulas, going back \cite{Skla59}, has drawn a lot of attention. The copula $C: [0,1]^d \rightarrow [0,1]$ of a random vector $\mathbf{X} = (X_0, \dots, X_{d-1})$ with $d \geq 2$ allows us to separate the effect of dependence from the effect of the marginal distribution, such that:
    \begin{equation*}
        \mathbb{P}\left\{ X_0 \leq x_0, \dots, X_{d-1} \leq x_{d-1} \right\} = C\left(\mathbb{P} {X_0 \leq x_0}, \dots, \mathbb{P}{X_{d-1} \leq x_{d-1} }\right),
    \end{equation*}
    where $(x_0, \dots, x_{d-1}) \in \mathbb{R}^d$. The main consequence of this identity is that the copula completely characterizes the stochastic dependence between the margins of $\mathbf{X}$.
    
    In other words, copulae allow us to model marginal distributions and dependence structure separately. Furthermore, motivated by Sklar's theorem, the problem of investigating stochastic dependence is reduced to the study of multivariate distribution functions under the unit hypercube $[0,1]^d$ with uniform margins. The theory of copulae has been of prime interest for many applied fields of science, such as quantitative finance (\cite{patton2012review}) or environmental sciences (\cite{MISHRA2011157}). This increasing number of applications has led to a demand for statistical methods. For example, semiparametric estimation (\cite{10.2307/2337532}), nonparametric estimation (\cite{10.2307/3318798}) of copulae or nonparametric estimation of conditional copulae (\cite{10.2307/24586878, PORTIER2018160}) have been investigated. These results are established for a fixed arbitrary dimension $d \geq 2$, but several investigations (e.g. \cite{10.2307/25463423, 10.1214/21-AOS2050}) are done for functional data for the tail copula, which captures dependence in the upper tail.
    
    Software implementation of copulas has been extensively studied in \textsf{R}, for example in the packages \cite{evdR, copulaR, VineCopulaR}. However, methods for working with copulas in \textsf{Python} are still limited. As far as we know, copula-dedicated packages in \textsf{Python} are mainly designed for modeling, such as \cite{copulasPy} and \cite{CopulaePy}. These packages use maximum likelihood methods to estimate the copula parameters from observed data and generate synthetic data using the estimated copula model. Other packages provide sampling methods for copulas, but they are typically restricted to the bivariate case and the conditional simulation method (see, for example, \cite{baudin2017openturns}). Additionally, these packages often only consider Archimedean and elliptical copulas, and do not include the extreme value class in arbitrary dimensions $d \geq 2$ (\cite{nicolas2022pycop}). In this paper, we propose to implement a wide range of copulas, including the extreme value class, in arbitrary fixed dimension $d \geq 2$.
    
    Through this paper we adopt the following notational conventions: all the indices will start at $0$ as in \textsf{Python}. Consider $(\Omega, \mathcal{A}, \mathbb{P})$ a probability space and let $\textbf{X} = (X_0, \dots, X_{d-1})$ be a $d$-dimensional random vector with values in $(\mathbb{R}^d, \mathcal{B}(\mathbb{R}^d))$, with $d \geq 2$ and $\mathcal{B}(\mathbb{R}^d)$ the Borel $\sigma$-algebra of $\mathbb{R}^d$. This random vector has a joint distribution $F$ with copula $C$ and its margins are denoted by $F_j(x) = \mathbb{P}\{X_j \leq x\}$ for all $x \in \mathbb{R}$ and $j \in \{0, \dots, d-1\}$. Denote by $\textbf{U} = (U_0, \dots, U_{d-1})$ a $d$ random vector with copula $C$ and uniform margins. All bold letters $\textbf{x}$ will denote a vector of $\mathbb{R}^d$.

    The \textsf{clayton} package, whose Python code can be found in this \href{https://github.com/Aleboul/clayton}{GitHub repository}, uses object-oriented features of the Python language. The package contains classes for Archimedean, elliptical, and extreme value copulas. In section \ref{sec:classes}, we briefly describe the classes defined in the package. Section \ref{sec:rng} presents methods for generating random vectors. In section \ref{sec:pairwise}, we apply the \textsf{clayton} package to model pairwise dependence between maxima. Section \ref{sec:discussion} discusses potential improvements to the package and provides concluding remarks. The appendices from \ref{app:bv_arch} to \ref{app:mv_ellip} define and illustrate all the parametric copula models implemented in the package.
    
    \section{Classes}
    \label{sec:classes}
        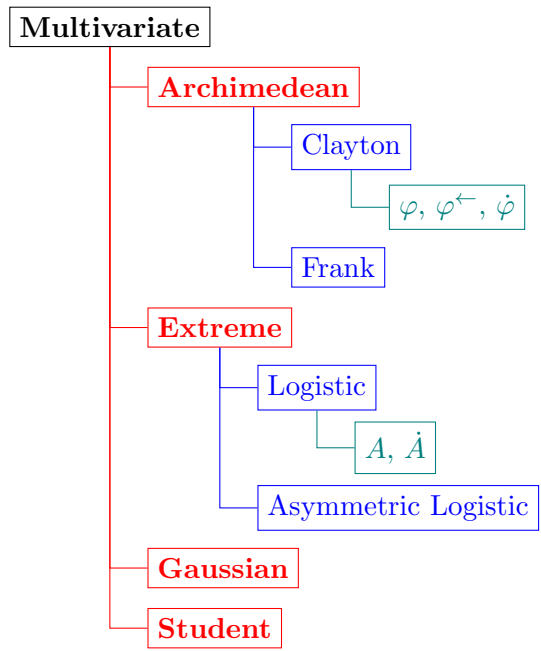
\begin{figure}
        \centering
            \begin{tikzpicture}
            [
                level 1/.style = {red},
                level 2/.style = {blue},
                level 3/.style = {teal},
                every node/.append style = {draw, anchor = west},
                grow via three points={one child at (0.5,-0.8) and two children at (0.5,-0.8) and (0.5,-1.6)},
                edge from parent path={(\tikzparentnode\tikzparentanchor) |- (\tikzchildnode\tikzchildanchor)}]
             
            \node {\textbf{Multivariate}}
                child {node {\textbf{Archimedean}}
                child {node {Clayton}
                child {node {$\varphi$, $\varphi^\leftarrow$, $\dot{\varphi}$}}}
                child [missing] {}
                child {node {Frank}}
                edge from parent }
                child [missing] {}
                child [missing] {}
                child [missing] {}
                child {node {\textbf{Extreme}}
                child {node {Logistic}
                child {node {$A$, $\dot{A}$}}}
                child [missing] {}
                child {node {Asymmetric Logistic}}
                edge from parent}
                child [missing] {}
                child [missing] {}
                child [missing] {}
                child {node {\textbf{Gaussian}}}
                child {node {\textbf{Student}}};
             
            \end{tikzpicture}
            \caption{The figure shows a object diagram that structures the code. The \textbf{Multivariate} class serves as the root and is used to instantiate all its child classes \textbf{Archimedean}, \textbf{Extreme}, \textbf{Gaussian}, and \textbf{Student} in red. The blue-colored classes correspond to various parametric copula models, and the green-colored classes represent examples of methods. Symbols $\varphi, \varphi^\leftarrow, \dot{\varphi}$ correspond to the generator function, its inverse, and its derivative, respectively, while $A, \dot{A}$ refer to the Pickands dependence function and its derivative.}
            \label{fig:diagram}
        \end{figure}
    	
    	The \textsf{clayton} package defines three main classes: \textbf{Multivariate}, \textbf{Archimedean}, and \textbf{Extreme}. The \textbf{Multivariate} class is designed for defining multivariate copulas (including the bivariate case) and is located at the highest level of the code architecture. It contains methods for instantiating a copula object or for sampling from a copula with desired margins using inversion methods, for example. The \textbf{Archimedean} and \textbf{Extreme} classes are children of the \textbf{Multivariate} class and represent copulas from the Archimedean and extreme value families, respectively. The \textbf{Gaussian} and \textbf{Student} classes represent elliptical copulas. This hierarchical structure is relevant theoretically, as Archimedean and extreme value copulas are studied independently (see, for example, \cite{charpentier2009} and \cite{genest2009rank}), and practically, as they contain effective sampling methods. However, the \textbf{Gaussian} and \textbf{Student} classes are split, as the most effective sampling algorithms are specific to each and cannot be generalized in a broader elliptical class.

        The architecture of the code is shown in Figure \ref{fig:diagram}. At the third level of the architecture, we find important parametric models of Archimedean and extreme value copulas (depicted as blue in the figure). These parametric models contain methods such as the generator function $\varphi$ (see Section \ref{subsec:Arch}) for Archimedean copulas and the Pickands dependence function $A$ (see Section \ref{subsec:Extreme}) for extreme value copulas (depicted as green in the figure). We provide a brief overview of Archimedean copulas and some of their properties in high-dimensional spaces in Section \ref{subsec:Arch}. A characterization of extreme value copulas is given in Section \ref{subsec:Extreme}. The appendices from \ref{app:bv_arch} to \ref{app:mv_ellip} define and illustrate all the copula models implemented in the package.

    	\subsection{The Archimedean class}
    	\label{subsec:Arch}
    	
    	Let $\varphi$ be a generator that is a strictly decreasing, convex function from $[0,1]$ to $[0, \infty]$ such that $\varphi(1) = 0$ and $\varphi(0) = \infty$. We denote the generalized inverse of $\varphi$ by $\varphi^\leftarrow$. Consider the following equation:
    	\begin{equation}
    		\label{eq:arch_cop}
    	 	C(\textbf{u}) = \varphi^\leftarrow (\varphi(u_0)+ \dots + \varphi(u_{d-1})).
    	\end{equation}
    If this relation holds and $C$ is a copula function, then $C$ is called an Archimedean copula. A necessary condition for \eqref{eq:arch_cop} to be a copula is that the generator $\varphi$ is a $d$-monotonic function, i.e., it is differentiable up to the order $d$ and its derivatives satisfy
    \begin{equation}
        \label{eq:dmono}
    		(-1)^k \left(\varphi\right)^{(k)}(x) \geq 0, \quad k \in \{1, \dots, d\}
    \end{equation}
    for $x \in (0, \infty)$ (see Corollary 2.1 of \cite{10.1214/07-AOS556}). Note that $d$-monotonic Archimedean inverse generators do not necessarily generate Archimedean copulas in dimensions higher than $d$ (see \cite{10.1214/07-AOS556}). As a result, some Archimedean subclasses are only implemented for the bivariate case as they do not generate an Archimedean copula in higher dimensions. In the bivariate case, \eqref{eq:dmono} can be interpreted as $\varphi$ being a convex function.

    The \textsf{clayton} package implements common one-parameter families of Archimedean copulas, such as the Clayton (\cite{10.2307/2335289}), Gumbel (\cite{1960}), Joe (\cite{joe1997multivariate}), Frank (\cite{Frank1979}), and AMH (\cite{ALI1978405}) copulas for the multivariate case. It is worth noting that all Archimedean copulas are symmetric, and in dimensions 3 or higher, only positive associations are allowed. For the specific bivariate case, the package also implements other families, such as those numbered from 4.2.9 to 4.2.15 and 4.2.22 in Section 4.2 of \cite{nelsen2007introduction}. Definitions and illustrations of these parametric copula models can be found in appendices \ref{app:bv_arch} and \ref{app:mv_arch}.

    	\subsection{The Extreme class}
    	\label{subsec:Extreme}
    	
    	Investigating the notion of copulae within the framework of multivariate extreme value theory leads to the extreme value copulae (see \cite{gudendorf2009extremevalue} for an overview) defined as 
        \begin{equation}
        	\label{eq:evc}
        	C(\textbf{u}) = \exp \left( - \ell(-\ln(u_0), \dots, -\ln(u_{d-1})) \right), \quad \textbf{u} \in (0,1]^d,
        \end{equation}
        where $\ell: [0,\infty)^d \rightarrow [0,\infty)$ the stable tail dependence function which is convex, homogeneous of order one, namely $\ell(c\textbf{x}) = c \ell(\textbf{x})$ for $c > 0$ and satisfies $\max(x_0,\dots,x_{d-1})  \leq \ell(x_0,\dots,x_{d-1}) \leq x_0+\dots+x_{d-1}, \forall \textbf{x} \in [0,\infty)^d$. Let $\Delta^{d-1} = \{\textbf{w} \in [0,1]^d: w_0 + \dots + w_{d-1} = 1\}$ be the unit simplex. The Pickands dependence function $A: \Delta^{d-1} \rightarrow [1/d,1]$ characterizes $\ell$ by its homogeneity, which is the restriction of $\ell$ to the unit simplex $\Delta^{d-1}$:
        \begin{equation}
        	\label{eq:tail_dependence_pickands}
        	\ell(x_0, \dots,x_{d-1}) = (x_0 + \dots + x_{d-1}) A(w_0, \dots, w_{d-1}), \quad w_j = \frac{x_j}{x_0 + \dots + x_{d-1}},
        \end{equation}
        for $j \in \{1,\dots,d-1\}$ and $w_0 = 1 - w_1 - \dots - w_{d-1}$ with $\textbf{x} \in [0, \infty)^d \setminus \{\textbf{0}\}$. The Pickands dependence function characterizes the extremal dependence structure of an extreme value random vector and verifies $\max\{w_0,\dots,w_{d-1}\} \leq A(w_0,\dots,w_{d-1}) \leq 1$ where the lower bound corresponds to comonotonicity and the upper bound corresponds to independence. Estimating this function is an active area of research, with many compelling studies having been conducted on the topic (see, for example, \cite{bucher2011new, gudendorf2009extremevalue}).
        
        From a practical point of view, the family of extreme value copulae is very rich and arises naturally as the limiting distribution of properly normalised componentwise maxima. Furthermore, it contains a rich variety of parametric models and allows asymmetric dependence, that is, for the bivariate case:
            \begin{equation*}
                C(u_0,u_1) \neq C(u_1,u_0).
            \end{equation*}
        In the multivariate framework, the logistic copula (or Gumbel, see \cite{1960}), the asymmetric logistic copula (\cite{tawn1990}), the Hüsler and Reiss distribution (\cite{HUSLER1989283}), the t-EV copula (\cite{Demarta_Mcneil}), Bilogistic model (\cite{Smith1990}) are implemented. It's worth noting that the logistic copula is the sole model that is both Archimedean and extreme value. The library includes bivariate extreme value copulae such as are asymmetric negative logistic (\cite{Joe1990FamiliesOM}), asymmetric mixed (\cite{10.1093/biomet/75.3.397}). The reader is again invited to read from \ref{app:bv_ext} to \ref{app:mv_ext} for precise definitions of these models.
	
	\section{Random vector generator}
	\label{sec:rng}
	
	We propose a \textsf{Python}-based implementation of a random vector generator that is capable of generating random vectors from a wide variety of copulas. The \textsf{clayton} package requires a few external libraries in order to function properly. These libraries are commonly used in scientific \textsf{Python} programming and are easy to install.

    The required libraries are:
    
    \begin{itemize}
    	\item \textsf{numpy} version 1.6.1 or newer. This is the fundamental package for scientific computing, it contains linear algebra functions and matrix / vector objects (\cite{harris2020array}).
    	\item \textsf{scipy} version 1.7.1 or newer. A library of open-source software for mathematics, science and engineering (\cite{virtanen2020scipy}).
    \end{itemize}

    The \textsf{clayton} package provides two methods for generating random vectors: \texttt{sample\_unimargin} and \texttt{sample}. The first method generates a sample where the margins are uniformly distributed on the unit interval $[0,1]$, while the second method generates a sample from the chosen margins.

    In Section \ref{subsec:biv_case}, we present an algorithm that uses the conditioning method to sample from a copula. This method is very general and can be used for any copula that is sufficiently smooth (see Equations \eqref{eq:cond_sim} and \eqref{eq:cond_dist_mv} below). However, the practical infeasibility of the algorithm in dimensions higher than $2$ and the computational intensity of numerical inversion call for more efficient ways to sample in higher dimensions. The purpose of Section \ref{subsec:mv_case} is to present such methods and to provide details on the methods used in the \textsf{clayton} package. In each section, we provide examples of code to illustrate how to instantiate a copula and how to sample with \textsf{clayton}.
	
	In the following sections, we will use \textsf{Python} code that assumes that the following packages have been loaded:

    \begin{minted}{python}
        >>> import clayton
        >>> from clayton.rng import base, evd, archimedean, monte_carlo
        >>> import numpy as np
        >>> import matplotlib.pyplot as plt
        >>> from scipy.stats import norm, expon
        >>> np.random.seed(42)
    \end{minted}
	\subsection{The bivariate case}
	\label{subsec:biv_case}
	
	In this subsection, we address the problem of generating a bivariate sample from a specified joint distribution with $d=2$. Suppose that we want to sample a bivariate random vector $\textbf{X}$ with copula $C$. In the case where the components are independent, the sampling procedure is straightforward: we can independently sample $X_0$ and $X_1$. However, in the general case where the copula is not the independence copula, this approach is not applicable.

    One solution to this problem is to use the conditioning method to sample from the copula. This method relies on the fact that given $(U_0, U_1)$ with copula $C$, the conditonal law of $U_1$ given $U_0$ is written as: 
    \begin{equation}
    	\label{eq:cond_sim}
    	c_{u_0}(u_1) \triangleq \mathbb{P}\left\{ U_1 \leq u_1 | U_0 = u_0 \right\} = \frac{\partial C(u_0,u_1)}{\partial u_0}.
    \end{equation}
    This allows us to first sample $U_0$ from a uniform distribution on the unit interval, and then to use the copula to generate $U_1$ given $U_0$. Finally, we can transform the resulting sample $(U_0, U_1)$ into the original space by applying the inverse marginal distributions $F_0^{-1}$ and $F_1^{-1}$ to $U_0$ and $U_1$ respectively. Thus, an algorithm for sampling bivariate copulas is given in Algorithm \ref{alg:1}. Algorithm \ref{alg:1} presents a procedure for generating a bivariate sample from a copula. The algorithm takes as input the length of the sample $n$, as well as the parameters of the copula ($\theta, \psi_1, \psi_2$). The output is a bivariate sample from the desired copula model, denoted ${(u_0^{(1)},u_1^{(1)}), \dots, (u_0^{(n)},u_1^{(n)})}$. This algorithm is applicable as long as the copula has a first partial derivative with respect to its first component.

\begin{algorithm}

\caption{Conditional sampling from a copula}

\begin{algorithmic}[1]
\State \textbf{Data}: sample's length $n$.
\State Parameter of the copula $\theta, \psi_1, \psi_2$.
\State \textbf{Result}: Bivariate sample from the desired copula model $\{(u_0^{(1)},u_1^{(1)}), \dots, (u_0^{(n)},u_1^{(n)})\}$.
\Procedure{sampling}{$n, \theta, \psi_1, \psi_2$}
    \State Generate two independent uniform random observations on the $[0,1]$ segment $u_0$ and $t_1$.
    \State Set $u_1 = c_{u_0}^\leftarrow(t_1)$ where $c_{u_0}^\leftarrow$ denotes the generalized inverse of $c_{u_0}$.
    \State The desired pair is $(u_0,u_1)$.
\EndProcedure

\end{algorithmic}
\label{alg:1}
\end{algorithm}

    For step 6 of the algorithm, we need to find $u_1 \in [0,1]$ such that $c_{u_0}(u_1) - t_1 = 0$ holds. This $u_1$ always exists because for every $u \in ]0,1[$, we have $0 \leq c_{u_0}(u) \leq 1$, and the function $u \mapsto c_{u_0}(u)$ is increasing (see Theorem 2.2.7 of \cite{nelsen2007introduction} for a proof). This step can be solved using the \textsf{brentq} function from the \textsf{scipy} package. A sufficient condition for a copula to have a first partial derivative with respect to its first component in the Archimedean and extreme value cases is that the generator $\varphi$ and the Pickands dependence function $A$ are continuously differentiable on $]0,1[$, respectively. In this case, the first partial derivatives of the copula are given by:
    \begin{align*}
    	&\frac{\partial C}{\partial u_0}(u_0,u_1) = \frac{\varphi'(u_0)}{\varphi'(C(u_0,u_1))}, \quad (u_0,u_1) \in ]0,1[^2, \\
    	&\frac{\partial C}{\partial u_0}(u_0,u_1) = \frac{C(u_0,u_1)}{u_0} \mu(t), \quad  (u_0,u_1) \in ]0,1[^2,
    \end{align*}
    where $t = \ln(u_1) / \ln(u_0u_1) \in (0,1)$ and $\mu(t) = A(t) - tA'(t)$. 
    
    We now have all the necessary theoretical tools to give details on how the $\textsf{clayton}$ package is designed. The file $\texttt{base.py}$ contains the $\textbf{Multivariate}$ class and the $\texttt{sample}$ method to generate random numbers from $\textbf{X}$ with copula $C$. To do so, we use the inversion method that is to sample from $\textbf{U}$ using Algorithm \ref{alg:1} and we compose the corresponding uniform margins by $F_j^\leftarrow$. Equation \eqref{eq:cond_sim} indicates that the sole knowledge of $A$ and $\varphi$ and their respective derivatives are needed in order to perform the sixth step of Algorithm \ref{alg:1}. For that purpose, $\texttt{cond\_sim}$ method located inside $\textbf{Archimedean}$ and $\textbf{Extreme}$ classes performs Algorithm \ref{alg:1}. Then each child of the bivariate $\textbf{Archimedean}$ (resp. $\textbf{Extreme}$) class is thus defined by its generator $\varphi$ (resp. $A$), it's derivative $\varphi'$ (resp. $A'$) and it's inverse $\varphi^\leftarrow$ as emphasized in greed in Figure \ref{fig:diagram}. Namely, we perform Algorithm \ref{alg:1} for the $\textbf{Archimedean}$ subclasses $\texttt{Frank}$, $\texttt{AMH}$, $\texttt{Clayton}$ (when $\theta < 0$ for the previous three), $\texttt{Nelsen\_9}$, $\texttt{Nelsen\_10}$, $\texttt{Nelsen\_11}$, $\texttt{Nelsen\_12}$, $\texttt{Nelsen\_13}$, $\texttt{Nelsen\_14}$, $\texttt{Nelsen\_15}$ and $\texttt{Nelsen\_22}$. For the $\textbf{Extreme}$ class, such algorithm is performed for the $\texttt{AsyNegLog}$ and $\texttt{AsyMix}$. For other models, faster algorithms are known and thus implemented, we refer to Section \ref{subsec:mv_case} for details.
	
\begin{figure}[!htp]
    	\centering
    	\includegraphics[scale = 0.75]{"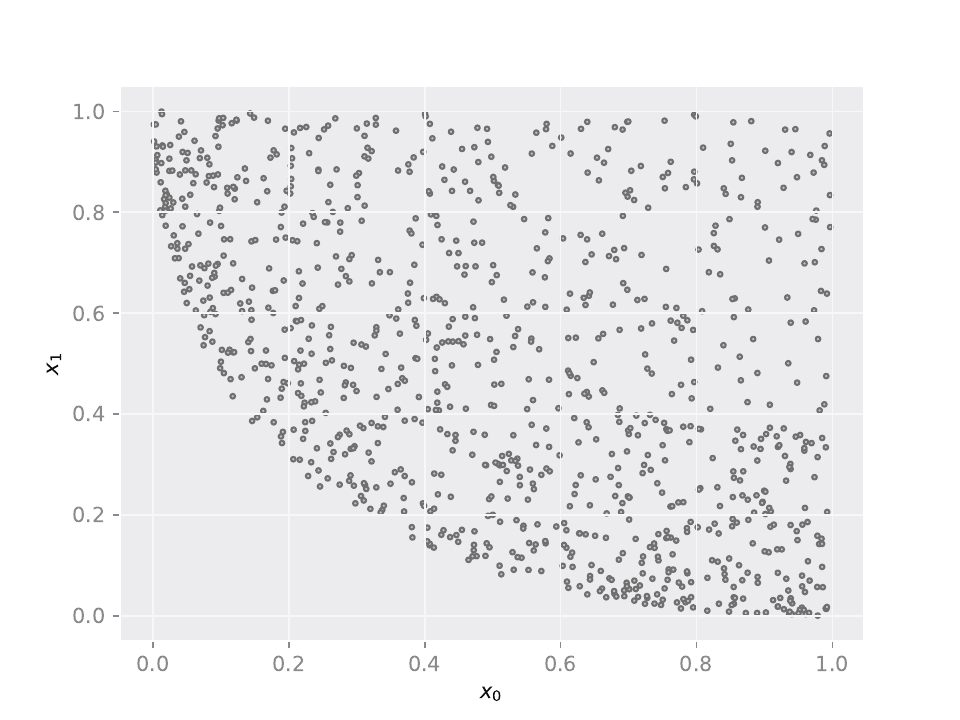"}
    	\label{fig:clayton_bv}
    	\caption{Scatterplot of a sample from a Clayton copula ($\theta = -0.5$).}
    	\label{fig:bv_plot}
    \end{figure}

	The following code illustrates the random vector generation for a bivariate Archimedean copula. By defining the parameter of the copula and the sample's length, the constructor for this copula is available and can be called using the \texttt{Clayton} method, such as:
	\begin{minted}{python}
	    >>> n_sample, theta = 1024, -0.5
	    >>> copula = archimedean.Clayton(theta=theta, n_sample=n_sample)
	\end{minted}
	To obtain a sample with uniform margins and a Clayton copula, we can use the \texttt{sample\_unimargin} method, as follows:
	\begin{minted}{python}
	    >>> sample = copula.sample_unimargin()
	\end{minted}
	Here, the \texttt{sample} object is a \textsf{numpy} array with $2$ columns and $1024$ rows, where each row contains a realization from a Clayton copula (see Figure \ref{fig:bv_plot}).
	
	\subsection{The multivariate case}
	\label{subsec:mv_case}
	We will now address the generation of multivariate Archimedean and Extreme value copulae proposed in the Clayton package. In the multivariate case, the link between partial derivatives and the conditional law remains. Indeed, let $(U_0, \dots, U_{d-1})$ be a $d$-dimensional random vector with uniform margins and copula $C$. The conditional distribution of $U_k$ given the values of $U_0, \dots, U_{k-1}$ is
    \begin{equation}
        \label{eq:cond_dist_mv}
        \mathbb{P}\left\{ U_k \leq u_k | U_0 = u_0, \dots, U_{k-1} = u_{k-1} \right\} = \frac{\partial^{k-1} C(u_0, \dots, u_k,1,\dots,1)/\partial u_0 \dots \partial u_{k-1}}{\partial^{k-1} C(u_0, \dots, u_{k-1},1,\dots,1) / \partial u_0 \dots \partial u_{k-1}},
    \end{equation}
    for $k \in {1,\dots, d-1}$. The conditional simulation algorithm may be written as follows.
    \begin{enumerate}
    	\item Generate $d$ independent uniform random on $[0,1]$ variates $v_0, \dots, v_{d-1}$.
    	\item Set $u_0 = v_0$.
    	\item For $k = 1, \dots, d-1$, evaluate the inverse of the conditional distribution given by Equation \eqref{eq:cond_dist_mv} at $v_k$, to generate $u_k$.
    \end{enumerate}
    Nevertheless, the evaluation of the inverse conditional distribution becomes increasingly complicated as the dimension $d$ increases. Furthermore, it can be difficult for some models to derive a closed form of Equation \eqref{eq:cond_dist_mv} that makes it impossible to implement it in a general algorithm with only the dimension $d$ as an input. For multivariate Archimedean copulas, \cite{10.1214/07-AOS556} give a method to generate a random vector from the $d$-dimensional copula $C$ with generator $\varphi$ (see Section 5.2 of \cite{10.1214/07-AOS556}). A stochastic representation for Archimedean copulas generated by a $d$-monotone generator is given by

    \begin{equation}
        \label{eq:radial}
        \textbf{U} = \left( \varphi^\leftarrow(R S_1), \dots, \varphi^\leftarrow(RS_d) \right) \sim C,
    \end{equation}
    where $R \sim F_R$, the radial distribution which is independent of $S$ and $S$ is distributed uniformly in the unit simplex $\Delta^{d-1}$. One challenging aspect of this algorithm is to have an accurate evaluation of the radial distribution of the Archimedean copula and thus to numerically inverse this distribution. The associated radial distribution for the \textsf{Clayton} copula is given in Example 3.3 \cite{10.1214/07-AOS556} while those of the \textsf{Joe}, \textsf{AMH}, \textsf{Gumbel} and \textsf{Frank} copulas are given in \cite{hofert2012likelihood}. In general, one can use numerical inversion algorithms for computing the inverse of the radial distribution, however it will lead to spurious numerical errors. Other algorithms exist when the generator is known to be the Laplace-Stieltjes transform, denoted as $\mathcal{LS}$, of some positive random variables (see \cite{10.2307/2289314, frees1998understanding}). This positive random variable is often referenced as the frailty distribution. In this framework, Archimedean copulas allow for the stochastic representation
    \begin{equation}
        \textbf{U} = \left( \varphi^\leftarrow (E_1/V), \dots, \varphi^\leftarrow(E_d /V)\right) \sim C,
    \end{equation}
    with $V \sim F = \mathcal{LS}^{-1}[\varphi^\leftarrow]$ the frailty and $E_1, \dots, E_d$ are distributed i.i.d. according to a standard exponential and independent of $V$. Algorithm \ref{alg:frailty} presents a procedure for generating a multivariate sample from an Archimedean copula where the frailty distribution is known. The algorithm takes as an input the length of the sample $n$, as well as the parameter of the copula $\theta$. The output is a $d$-variate sample from the desired copula model, denoted $\{(u_0^{(1)}, \dots, u_{d-1}^{(1)}), \dots, (u_0^{(n)},\dots,u_{d-1}^{(n)})$.

    \begin{algorithm}
    
    \caption{Sampling from Archimedean copula using frailty distribution}
    
    \begin{algorithmic}[1]
    \State \textbf{Data}: sample's length $n$.
    \State Parameter of the copula $\theta$.
    \State \textbf{Result}: multivariate sample from the desired copula model.
    \Procedure{sampling}{$n, \theta$}
        \State Sample $V \sim F = \mathcal{LS}^{-1}[\varphi^\leftarrow]$.
        \State Sample $E_1, \dots, E_d \overset{i.i.d.}{\sim} \mathcal{E}(1)$, independent of $V$. 
        \State Return $\textbf{U} = (\varphi^\leftarrow(E_1/V), \dots, \varphi^\leftarrow(E_d / V))$.
    \EndProcedure
    
    \end{algorithmic}
    \label{alg:frailty}
    \end{algorithm}

    In this framework, we define \texttt{\_frailty\_sim} method defined inside the \textbf{Archimedean} class which performs Algorithm \ref{alg:frailty}. Then, each Archimedean copula is defined by the generator $\varphi$, it's inverse $\varphi^\leftarrow$ and the frailty distribution denoted as $\mathcal{LS}^{-1}[\varphi^\leftarrow]$ as long as we know the frailty. This is the case for \texttt{Joe}, \texttt{Clayton}, \texttt{AMH} or \texttt{Frank}.

    For the extreme value case, algorithms have been proposed, as in \cite{stephenson2003simulating} (see Algorithms 2.1 and 2.2), who proposes sampling methods for the Gumbel and the asymmetric logistic model. These algorithms are implemented in the \textsf{clayton} package. Note that these algorithms are model-specific, thus the \texttt{sample\_unimargin} method is exceptionally located in the corresponding child of the multivariate \textbf{Extreme} class. Another procedure designed by \cite{10.1093/biomet/asw008} to sample from multivariate extreme value models using extremal functions (see Algorithm 2 of the reference cited above) is also of prime interest. For the implemented models using this algorithm, namely \textbf{Hüsler-Reiss}, \textbf{tEV}, \textbf{Bilogistic} and \textbf{Dirichlet} models, a method called \texttt{\_rextfunc} is located inside each classes which allows to generate an observation from the according law of the extremal function.

    Samples from the Gaussian and Student copula are directly given by Algorithm 5.9 and 5.10 respectively of \cite{quantrisk}. As each algorithm is model specific, the \texttt{sample\_unimargin} method is located inside the \textbf{Gaussian} and \textbf{Student} classes.

\begin{figure}[!htp]
    	\centering
    	\includegraphics[scale = 0.5]{"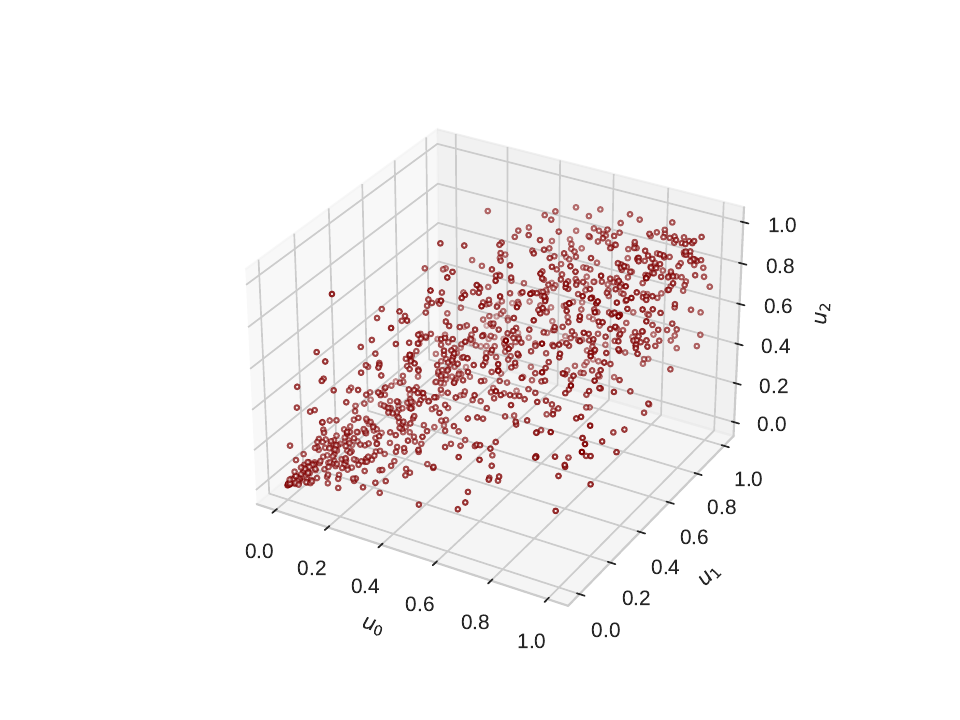"}
    	\caption{Scatterplot of a sample from a Clayton copula ($\theta = 2.0$).}
    	\label{fig:mv_plot}
    \end{figure}

    We present how to construct a multivariate Archimedean copula and to generate random vectors from this model. Introducing the parameters of the copula, we appeal the following lines to construct our copula object:
    \begin{minted}{python}
        >>> d, theta, n_sample = 3, 2.0, 1024
        >>> copula = archimedean.Clayton(theta=theta, n_sample=n_sample,
        >>> dim=d)
    \end{minted}
    We now call the \texttt{sample\_unimargin} method to obtain randomly generated vectors.
    \begin{minted}{python}
        sample = copula.sample_unimargin()
    \end{minted}
    We thus represent in three dimensions in Figure \ref{fig:mv_plot}.
    
    \section{Case study: modeling pairwise dependence between spatial maximas with missing data}
    \label{sec:pairwise}
    
    We now proceed to a case study where we use our \textsf{Python} package to assess, under a finite sample framework, the asymptotic properties of an estimator of the $\lambda$-madogram when data are completely missing at random (MCAR). This case study comes from numerical results of \cite{boulin2021non}. The $\lambda$-madogram belongs to a family of estimators, namely the madogram, which is of prime interest in environmental sciences, as it is designed to model pairwise dependence between maxima in space. See, for example, \cite{bernard:hal-03207469,BADOR201517,saunders}, where the madogram was used as a dissimilarity measure to perform clustering. In several fields, such as econometrics (\cite{woolridge2007}) or survey theory (\cite{chauvet2015}), the MCAR hypothesis appears to be a strong hypothesis, but in environmental research, this hypothesis is more realistic, as the missingness of one observation is usually due to instruments, communication, and processing errors that may be reasonably supposed to be independent of the quantity of interest. In Section \ref{subsec:background}, we define objects and properties of interest, while in Section \ref{subsec:num}, we describe a detailed tutorial in \textsf{Python} and with the \textsf{clayton} package to compare the asymptotic variance with an empirical counterpart of the $\lambda$-madogram with $\lambda = 0.5$.
    
    \subsection{Background}
    \label{subsec:background}
        
        It was emphasized that the possible dependence between maxima can be described with the extreme value copula. This function is completely characterized by the Pickands dependence function (see Equation \eqref{eq:tail_dependence_pickands}), which is equivalent to the $\lambda$-madogram introduced by \cite{naveau:hal-00312758} and defined as
        \begin{equation}
            \label{eq:lmbd_mado}
            \nu(\lambda) = \mathbb{E}\left[ \left|{F_0(X_0)}^{1/\lambda} - {F_1(X_1)}^{1/(1-\lambda)} \right|\right],
        \end{equation}
        with $\lambda \in (0,1)$, and if $\lambda = 0$ and $0<u<1$, then $u^{1/\lambda} = 0$ by convention. The $\lambda$-madogram took its inspiration from the extensively used geostatistics tool, the variogram (see Chapter 1.3 of \cite{alma991005826659705596} for a definition and some classical properties). The $\lambda$-madogram can be interpreted as the $L_1$-distance between the uniform margins elevated to the inverse of the corresponding weights $\lambda$ and $1-\lambda$. This quantity describes the dependence structure between extremes by its relation with the Pickands dependence function. If we suppose that $C$ is an extreme value copula as in Equation \eqref{eq:evc}, we have
        \begin{equation}
        \label{eq:pickands_mado}
            A(\lambda) = \frac{\nu(\lambda) + c(\lambda)}{1-\nu(\lambda) - c(\lambda)},
        \end{equation}
        with $c(\lambda) = 2^{-1} (\lambda / (1-\lambda) + (1-\lambda)/\lambda)$ (see Proposition 3 of \cite{MARCON20171} for details).
        
        We consider independent and identically distributed i.i.d. copies $\textbf{X}_1, \dots, \textbf{X}_n$ of $\textbf{X}$. In the presence of missing data, we do not observe a complete vector $\textbf{X}_i$ for $i \in \{1,\dots,n\}$. We introduce $\textbf{I}_i \in \{0,1\}^2$ which satisfies, $\forall j \in \{0,1\}$, $I_{i,j} = 0$ if $X_{i,j}$ is not observed. To formalize incomplete observations, we introduce the incomplete vector $\tilde{\textbf{X}}_i$ with values in the product space $\bigotimes_{j=1}^2 (\mathbb{R} \cup \{\textsf{NA}\})$ such as
        \begin{equation*}
        	\tilde{X}_{i,j} = X_{i,j} I_{i,j} + \textsf{NA} (1-I_{i,j}), \quad i \in \{1,\dots,n\}, \, j \in \{0,\dots, d-1\}.
        \end{equation*}
        We thus suppose that we observe a $4$-tuple such as
        \begin{equation}
        \label{missing_2}
        	(\textbf{I}_i, \tilde{\textbf{X}}_i), \quad i \in \{1,\dots,n\},
        \end{equation}
        i.e. at each $i \in \{1,\dots,n\}$, several entries may be missing. We also suppose that for all $i \in \{1, \dots,n \}$, $\textbf{I}_{i}$ are i.i.d copies from $\textbf{I} = (I_0, I_1)$ where $I_j$ is distributed according to a Bernoulli random variable $\mathcal{B}(p_j)$ with $p_j = \mathbb{P}(I_j = 1)$ for $j \in \{0,1\}$. We denote by $p$ the probability of observing completely a realization from $\textbf{X}$, that is $p = \mathbb{P}(I_0=1, I_1 = 1)$. In \cite{boulin2021non}, hybrid and corrected estimators, respectively denoted as $\hat{\nu}_n^{\mathcal{H}}$ and $\hat{\nu}_n^{\mathcal{H*}}$, are proposed to estimate nonparametrically the $\lambda$-madogram in presence of missing data completely at random. Furthermore, a closed expression of their asymptotic variances for $\lambda \in ]0,1[$ is also given. This result is summarized in the following proposition.
        \begin{proposition}[\cite{boulin2021non}]
        	\label{prop:boulin}
        	Let $(\textbf{I}_i, \tilde{\textbf{X}_i})_{i=1}^n$ be a sample given by \eqref{missing_2}. For $\lambda \in ]0,1[$, if $C$ is an extreme value copula in \eqref{eq:evc} with Pickands dependence function $A$, we have
                        \begin{align*}
                            &\mathcal{E}_n^{\mathcal{H}}(\lambda) \triangleq \sqrt{n} \left(\hat{\nu}_n^{\mathcal{H}}(\lambda) - \nu( \lambda)\right) \overunderset{d}{n \rightarrow \infty}{\rightarrow} \mathcal{N}\left(0, \mathcal{S}^{\mathcal{H}}(p_0,p_1,p, \lambda)\right), \\
                            &\mathcal{E}_n^{\mathcal{H}*}(\lambda) \triangleq \sqrt{n} \left(\hat{\nu}_n^{\mathcal{H}*}(\lambda) - \nu( \lambda)\right) \overunderset{d}{n \rightarrow \infty}{\rightarrow} \mathcal{N}\left(0, \mathcal{S}^{\mathcal{H}*}(p_0,p_1,p, \lambda)\right),
                        \end{align*}
                        where $\nu(\lambda)$ is defined in \eqref{eq:lmbd_mado}, $\mathcal{S}^{\mathcal{H}}(p_0,p_1,p, \lambda)$ and $\mathcal{S}^{\mathcal{H}*}(p_0,p_1,p, \lambda)$ are the asymptotic variances of the random variables where the closed expression is given in \cite{boulin2021non}, Proposition 1.
        \end{proposition}

    \subsection{Numerical results}
    \label{subsec:num}
    
\begin{figure}[!htp]
        \centering
        \includegraphics[scale=0.75]{"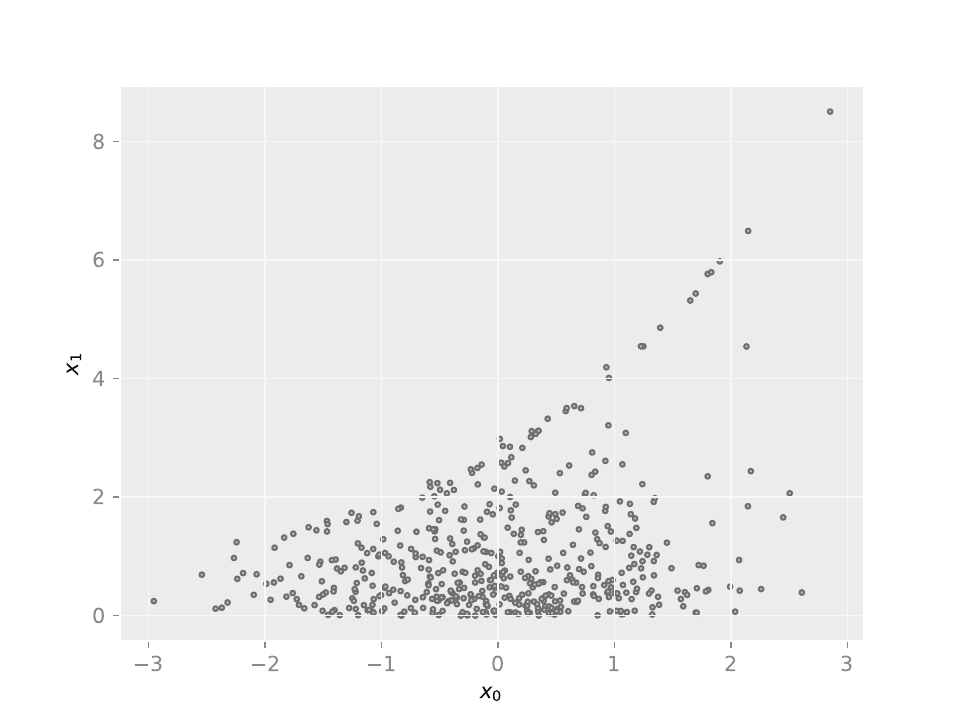"}
        \caption{A realization from the \textbf{asymmetric negative logistic model} with Gaussian and Exponent margins and parameters $\theta = 10, \psi_1 = 0.1, \psi_2 = 1.0$ and sample's length $n = 1024$.}
        \label{fig:plot_margins}
    \end{figure}

\begin{figure}[!ht]
        \centering
        \includegraphics[scale = 0.75]{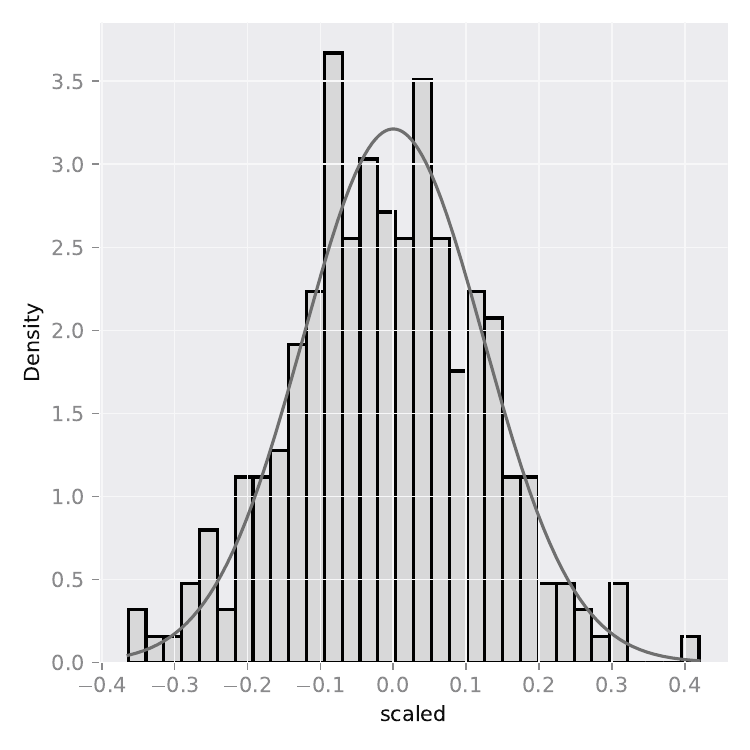}
        \caption{Histogram of $\mathcal{E}_n^{\mathcal{H}*}$ in Proposition \ref{prop:boulin} where the solid line is the density of a centered Gaussian with variance $\mathcal{S}^{\mathcal{H}*}$.}
        \label{fig:bv_normality}
    \end{figure}

    Benefiting from generating data with \textsf{clayton} we are thus able, with Monte Carlo simulation, to assess theoretical results given by Proposition \ref{prop:boulin} in a finite sample setting. For that purpose, we implement a \textsf{MonteCarlo} class (in \texttt{monte\_carlo.py} file) which contains some methods to perform some Monte Carlo iterations for a given extreme value copula. Now, we set up parameters to sample our bivariate dataset. For this subsection, we choose the asymmetric negative logistic model (see \ref{app:bv_ext} for a definition) with parameters $\theta = 10, \psi_1 = 0.1, \psi_2 = 1.0$. 
    \begin{minted}{python}
        >>> n_sample = 1024
        >>> theta, psi1, psi2 = 10, 0.1, 1.0
    \end{minted}
    We choose the standard normal and exponential as margins. To simulate this sample, the following lines should be typed:
    \begin{minted}{python}
        >>> copula = evd.AsyNegLog(theta=theta, psi1=psi1, psi2=psi2,
        >>> n_sample=n_sample)
        >>> sample = copula.sample(inv_cdf=[norm.ppf, expon.ppf])
    \end{minted}
    The $1024 \times 2$ array \texttt{sample} contains $1024$ realization of the \textbf{asymmetric negative logistic} model where the first column is distributed according to a standard normal random variable and the second column as a standard exponential. This distribution is depicted in Figure \ref{fig:plot_margins}. To obtain it, one needs the following lines of command:
    \begin{minted}{python}
        >>> fig, ax = plt.subplots()
        >>> ax.scatter(sample[:,0], sample[:,1],
        >>> edgecolors='#6F6F6F', color='#C5C5C5', s=5)
        >>> ax.set_xlabel(r'$x_0$')
        >>> ax.set_ylabel(r'$x_1$')
        >>> plt.show()
    \end{minted}
    Before going into further details, we will present the missing mechanism. Let $V_0$ and $V_1$ be random variables uniformly distributed under the $]0,1[$ segment with copula $C_{(V_0,V_1)}$. We set $I_0 = \mathds{1}{{V_0 \leq p_0}}$ and $I_1 = \mathds{1}{{V_1 \leq p_1}}$. It is thus immediate that $I_0 \sim \mathcal{B}(p_0)$ and $I_1 \sim \mathcal{B}(p_1)$ and $p \triangleq \mathbb{P}{I_0 = 1, I_1 =1 } = C_{(V_0,V_1)}(p_0, p_1)$. For our illustration, we will take $C_{(V_0,V_1)}$ as a \texttt{Joe} copula with parameter $\theta = 2.0$ (we refer to \ref{app:bv_arch} for a definition of this copula). For this copula, it is more likely to observe a realization $v_0 \geq 0.8$ from $V_0$ if $v_1 \geq 0.8$ from $V_1$. If we observe $v_1 < 0.8$, the realization $v_0$ is close to being independent of $v_1$. In climate studies, extreme events could damage the recording instrument in the surrounding regions where they occur, thus the missingness of one variable may depend on others. We initialize the copula $C_{(V_0,V_1)}$ with the following line:
    \begin{minted}{python}
        >>> copula_miss = archimedean.Joe(theta=2.0, n_sample=n_sample)
    \end{minted}
    For a given $\lambda \in ]0,1[$, we now want to estimate a $\lambda$-madogram with a sample from the asymmetric negative logistic model, where some observations are missing due to the missing mechanism described above. We will repeat this step several times to compute an empirical counterpart of the asymptotic variance. The \texttt{MonteCarlo} object has been designed for this purpose: we specify the number of iterations $n_{iter}$ (take $n_{iter} = 1024$), the chosen extreme value copula (asymmetric negative logistic model), the missing mechanism (described by $C_{(V_0,V_1)}$ and $p_0 = p_1 = 0.9$), and $\lambda$ (noted \texttt{w}). We can write the following lines of code:
    \begin{minted}{python}
        >>> u = np.array([0.9, 0.9])
        >>> n_iter, P, w = 256, [[u[0], copula_miss._c(
        >>>     u)], [copula_miss._c(u), u[1]]], np.array([0.5, 0.5])
        >>> monte = monte_carlo.MonteCarlo(n_iter=n_iter, n_sample=n_sample,
        >>>     copula=copula, copula_miss=copula_miss, weight=w, matp=P)
    \end{minted}
    The \texttt{MonteCarlo} object is thus initialized with all parameters needed. We may use the \texttt{simu} method to generate a \texttt{DataFrame} (a \texttt{Pandas} object) composed out $1024$ rows and $3$ columns. Each row contains an estimate of the $\lambda$-madogram, $\hat{\nu}_n^{\mathcal{H}*}$ in Proposition \ref{prop:boulin} (\texttt{FMado}), the sample length $n$ (\texttt{n}) and the normalized estimation error (\texttt{scaled}). We thus call the \texttt{simu} method.
    \begin{minted}{python}
    >>> df_wmado = monte.simu(inv_cdf = [norm.ppf, expon.ppf], corr = True)
    >>> print(df_wmado.head())
    	     FMado		n		    scaled
    0	 0.147648	   512.0		 -0.140255
    1	 0.160095	   512.0		 -0.141402
    2	 0.159303	   512.0		  0.123480
    3	 0.156156	   512.0 		 0.052269
    4	 0.152242	   512.0		 -0.036300
    \end{minted}
    Where \texttt{corr=True} specifies that we compute the corrected estimator, $\hat{\nu}_n^{\mathcal{H}*}$ in Proposition \ref{prop:boulin}. Now, using the \texttt{var\_mado} method defined inside in the \textbf{Extreme} class, we obtain the asymptotic variance for the given model and parameters from the missing mechanism. We obtain this quantity as follows
    \begin{minted}{python}
    >>> var_mado = copula.var_mado(w, p=copula_miss._c(u), P=P, corr=True)
    >>> print(var_mado)
    0.015417245591834503
    \end{minted}
    We propose here to check numerically the asymptotic normality with variance $\mathcal{S}^{\mathcal{H}*}$ of the normalized estimation error of the corrected estimator. We have all data in hand and the asymptotic variance was computed by lines above. We thus write:
    \begin{minted}{python}
    >>> fig, ax = plt.subplots()
    >>> sigma = np.sqrt(var_mado)
    >>> x = np.linspace(min(df_wmado['scaled']), max(df_wmado['scaled']), 1000)
    >>> gauss = gauss_function(x, 0, sigma)
    >>> sns.displot(data=df_wmado, x="scaled", color='#C5C5C5', kind='hist',
    >>>     stat='density', common_norm=False, alpha=0.5, fill=True,
    >>>      linewidth=1.5, bins = 32)
    >>> plt.plot(x,gauss, color = 'darkblue')
    >>> plt.show()
    \end{minted}
    Result of these lines might be found in Figure \ref{fig:bv_normality}.
    \section{Discussion}
    \label{sec:discussion}
    \subsection{Comparison of \textsf{clayton} with \textsf{R} packages}

    To compare \textsf{clayton} to existing packages in \textsf{R}, we consider the \textsf{copula} package (\cite{kojadinovic2010modeling}) and \textsf{mev} (\cite{mevR}) for sampling from Archimedean and multivariate extreme value distributions, respectively. To run the experiment, we use two computer clusters. The first cluster consists of five nodes, each with two 18-core Xeon Gold 3.1 GHz processors and 192 GB of memory, with 2933 MHz per socket. The second cluster has two CPU sockets, each containing a Xeon Platinum 8268 2.90 GHz processor with 24 cores. These configurations provide a significant amount of computational power and are well-suited for handling complex, data-intensive tasks. We use the first cluster to install the \textsf{copula} package and sample from the \textbf{Clayton}, \textbf{Frank}, and \textbf{Joe} models. We consider an increasing dimension $d \in \{50, 100, \dots, 1600\}$ for a fixed sample size of $n=1000$. We use the second cluster to install the \textsf{mev} package and call some of its methods to sample from the \textbf{Husler Reiss}, \textbf{Logistic}, and \textbf{TEV} distributions. Sampling from the latter is fast, but sampling from the two others is time consuming. Therefore, we only consider dimensions $d \in \{25, 50, \dots, 250\}$ for a fixed sample size of $n=1000$.

    \begin{figure}[!h]%
        \centering
        \subfloat[\centering Archimedean]{{\includegraphics[width=0.45\linewidth]{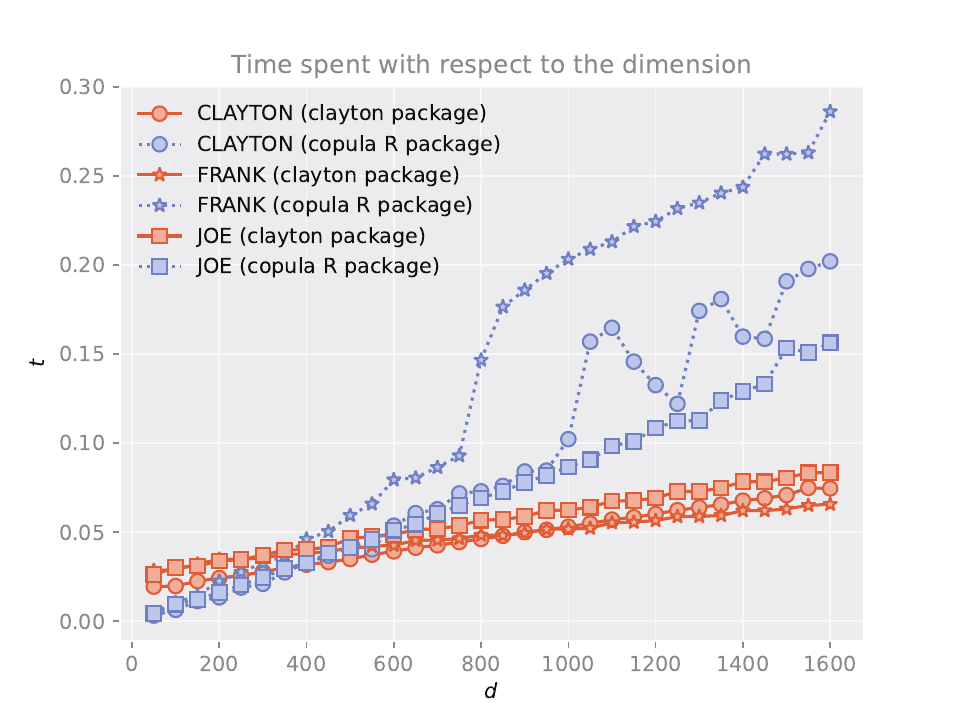} }}%
        \qquad
        \subfloat[\centering Multivariate extreme value]{{\includegraphics[width=0.45\linewidth]{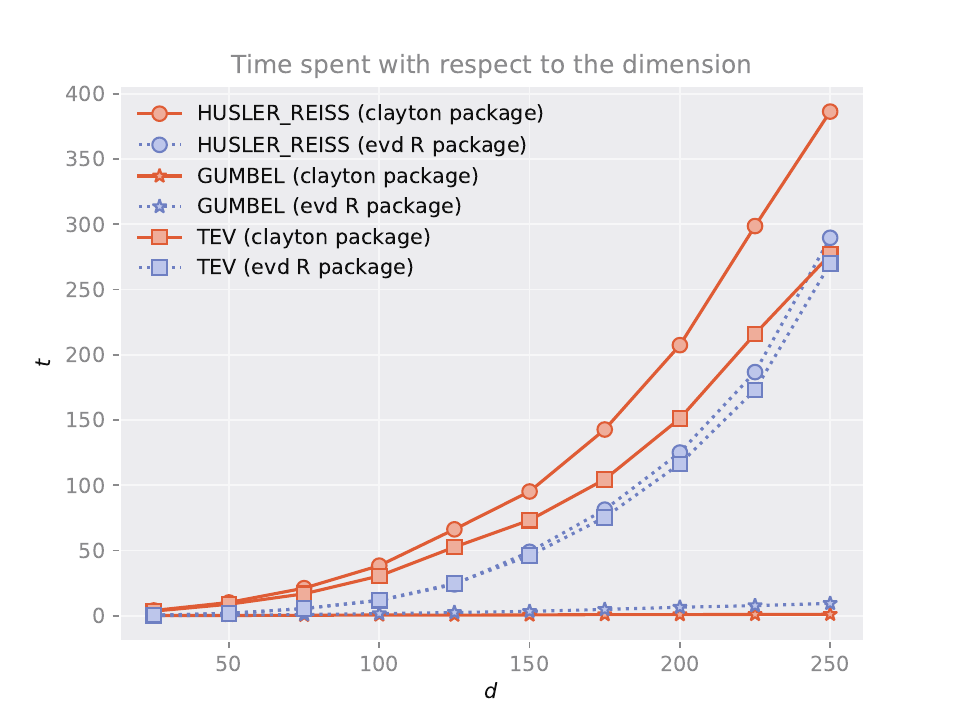} }}%
        \caption{Comparison results. Time spent (in seconds) to sample from the corresponding models with respect to the dimension $d$. The left panel shows the results for sampling from \textbf{Clayton}, \textbf{Frank} and \textbf{Joe} using \textsf{clayton} in \textsf{Python} and \textsf{copula} in \textsf{R}. The right panel shows the results for sampling from \textbf{HuslerReiss}, \textbf{Logistic} and \textbf{TEV} by \textsf{clayton} in \textsf{Python} and \textsf{mev} in \textsf{R}. In both cases, $1000$ vectors are generated for each model.}%
        \label{fig:com_num}%
    \end{figure}

    The figure shows the results of a comparison between the \textsf{clayton} and \textsf{copula} packages in \textsf{R}, and the \textsf{mev} package in \textsf{Python}. The comparison shows that the \textsf{clayton} package is more efficient at sampling from \textbf{Clayton}, \textbf{Frank} and \textbf{Joe} copulae than the \textsf{copula} package. The gap in efficiency may be due to the choice of algorithms used in the \textsf{clayton} package, which uses frailty distributions. The time required for sampling increases linearly with the dimension for the \textsf{clayton} package, but shows a more erratic behavior for the \textsf{copula} package.

    When comparing the \textsf{clayton} and \textsf{mev} packages, it is clear that \textsf{mev} is more efficient. This is likely due to the fact that \textsf{mev} is written in \textsf{C++}, while \textsf{clayton} is written in \textsf{Python}. The \textsf{mev} package uses the algorithm of \cite{stephenson2003simulating} to sample from the Logistic distribution, which is more efficient than the algorithm using frailty distributions used in \textsf{clayton}.
    
    \subsection{Conclusion}
    This paper presents the construction and some implementations of the \textsf{Python} package \textsf{clayton} for random copula sampling. This is a seminal work in the field of software implementation of copula modeling in \textsf{Python} and there is much more potential for growth. It is hoped that the potential diffusion of the software through those who need it may bring further implementations for multivariate modeling with copulas under \textsf{Python}. For example, choosing a copula to fit the data is an important but difficult problem. A robust approach to estimating copulas has been investigated recently by \cite{alquier2020estimation} using Maximum Mean Discrepancy. In relation to our example, semiparametric estimation of copulas with missing data could be of great interest, as proposed by \cite{HAMORI201985}.

    Additionally, implementation of the algorithm proposed by \cite{10.1214/07-AOS556} for generating random vectors for Archimedean copulas has been tackled, but as expected, numerical inversion gives spurious results, especially when the parameter $\theta$ and the dimension $d$ are high. Furthermore, as the support of the radial distribution is contained in the real line, numerical inversion leads to increased computational time. Further investigation is needed in order to generate random vectors from classical Archimedan models using the radial distribution.

    A direction of improvement for the \textsf{clayton} package is dependence modeling with Vine copulas, which have recently been a tool of high interest in the machine learning community (see, e.g., \cite{lopez2013gaussian, Veeramachaneni2015CopulaGM, Carrera2016VineCC, 10.5555/2946645.2946678} or \cite{ SunCuesta-InfanteVeeramachaneni2019}). This highlights the need for dependence modeling with copulas in \textsf{Python}, as a significant part of the machine learning community uses this language. In relation to this paper, Vine copulas may be useful for modeling dependencies between extreme events, as suggested by \cite{SIMPSON2021104736, nolde2021linking}. Furthermore, other copula models could be implemented to model further dependencies. These implementations will expand the scope of dependence modeling with \textsf{Python} and provide high-quality, usable tools for anyone who needs them.

\printbibliography

\appendix

\section{Bivariate archimedean models} \label{app:bv_arch}

\begin{table}[!htp]
\centering
\caption{Bivariate archimedean models in \textbf{COPPY} module.} \label{tab:arch_model_1}
  \begin{tabular}
      {ccccc} \hline Name & $\varphi(t)$ & Constraints & Figure \\
      
      \hline \textbf{Clayton} & $\frac{1}{\theta}(t^{-\theta}-1)$ &
      $\theta \in [-1, \infty) \setminus \{0\}$ & 
      \begin{minipage}{.3\textwidth}
        \centering
          \begin{tikzpicture}
                \draw (0, 0) node[inner sep=0] {\includegraphics[width=1.5in]{"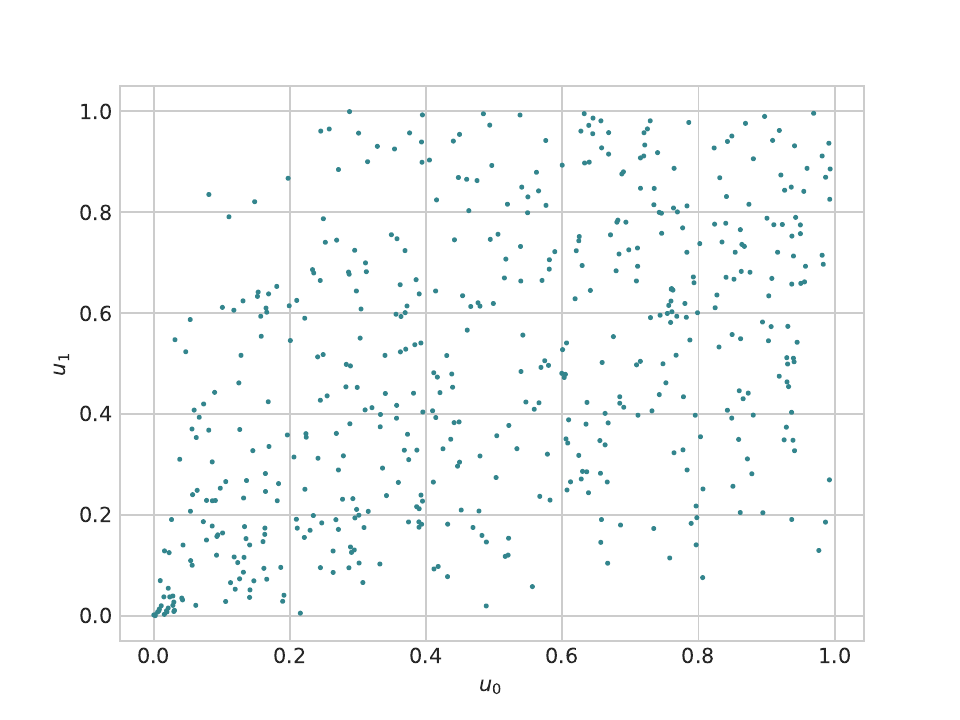"}};
                \draw (1, 1) node {\tiny$\theta = 1$};
            \end{tikzpicture} 
        \end{minipage}
        \\
      \textbf{AMH} & $\ln(\frac{1-\theta(1-t)}{t})$ & $\theta \in [-1, 1)$ & 
      \begin{minipage}{.3\textwidth}
        \centering
        \begin{tikzpicture}
            \draw (0, 0) node[inner sep=0] {\includegraphics[width=1.5in]{"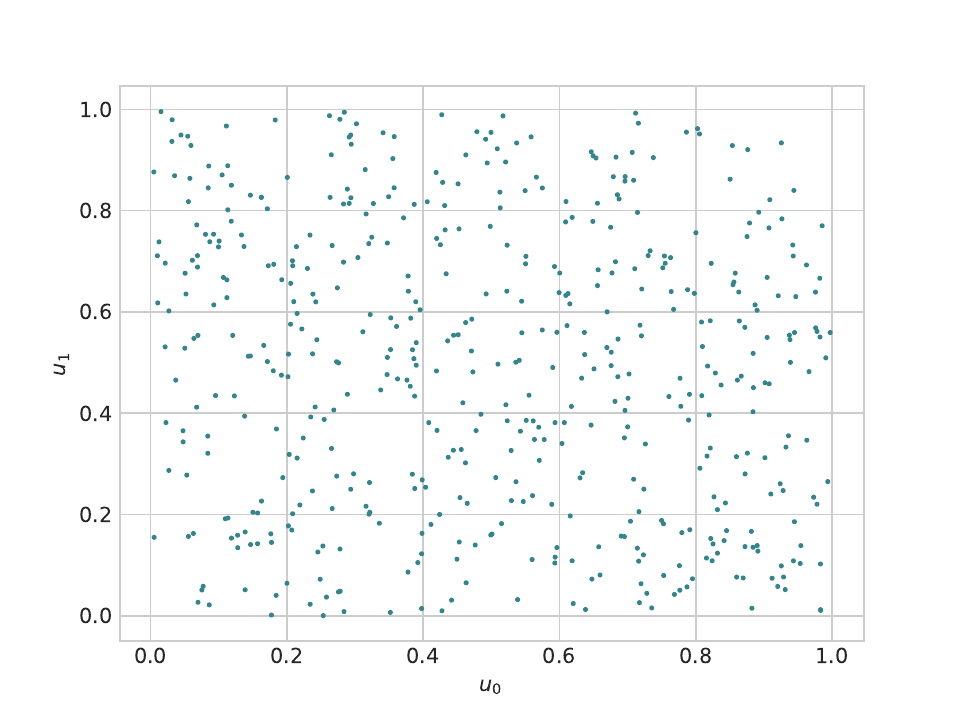"}};
            \draw (1, 1) node {\tiny$\theta = -0.5$};
        \end{tikzpicture}
        \end{minipage}
        \\
      \textbf{Frank} & $-\ln(\frac{e^{-\theta t}-1}{e^{-\theta}-1})$ & $\theta \in \mathbb{R}\setminus \{0\}$ &
      \begin{minipage}{.3\textwidth}
        \centering
        \begin{tikzpicture}
            \draw (0, 0) node[inner sep=0] {\includegraphics[width=1.5in]{"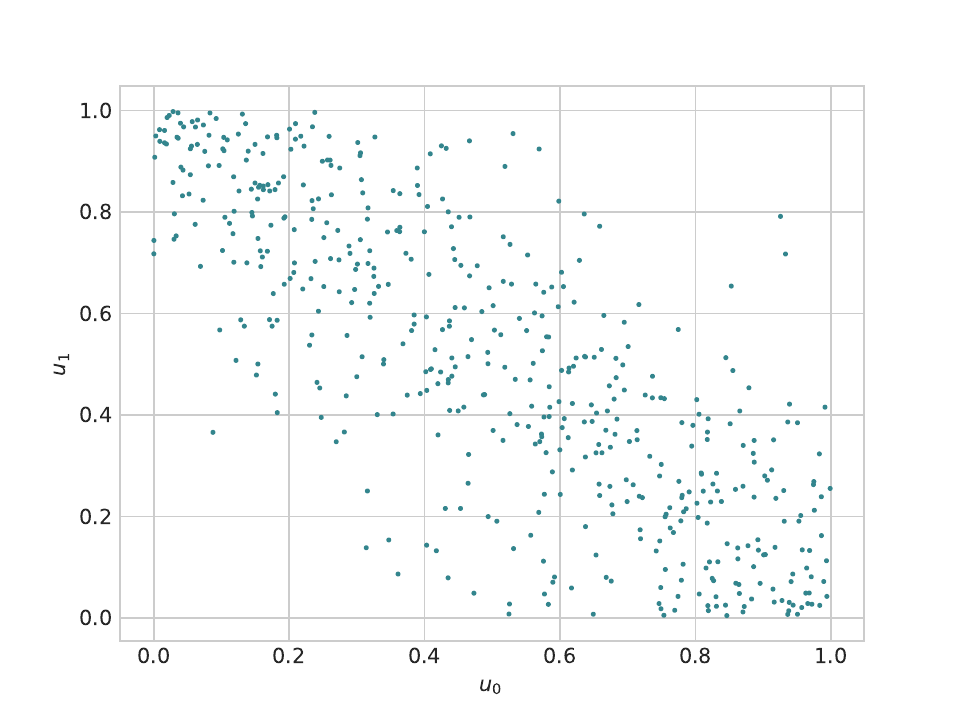"}};
            \draw (1, 1) node {\tiny$\theta = -8$};
        \end{tikzpicture}
        \end{minipage}
      \\
      \textbf{Joe} & $-\ln(1-(1-t)^\theta)$ & $\theta \in [1, \infty)$ &
      \begin{minipage}{.3\textwidth}
        \centering
        \begin{tikzpicture}
            \draw (0, 0) node[inner sep=0] {\includegraphics[width=1.5in]{"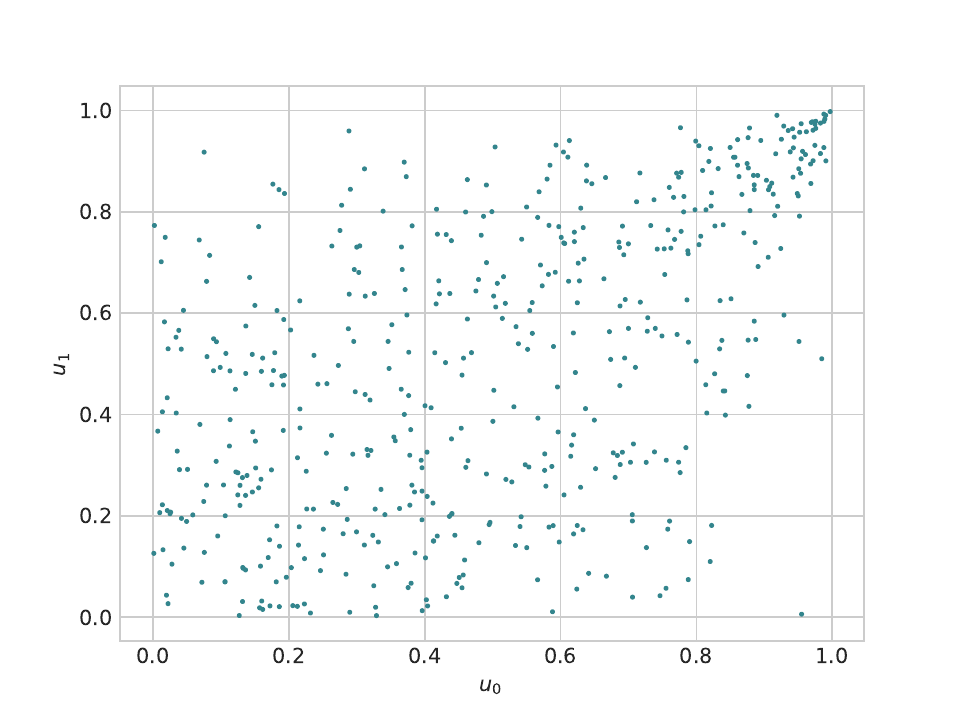"}};
            \draw (1, 1) node {\tiny$\theta = 2$};
        \end{tikzpicture}
        \end{minipage}
      \\
      \textbf{Nelsen n°9} & $\ln(1-\theta \ln(t))$ & $\theta \in ]0, 1]$ & 
      \begin{minipage}{.3\textwidth}
        \centering
        \begin{tikzpicture}
            \draw (0, 0) node[inner sep=0] {\includegraphics[width=1.5in]{"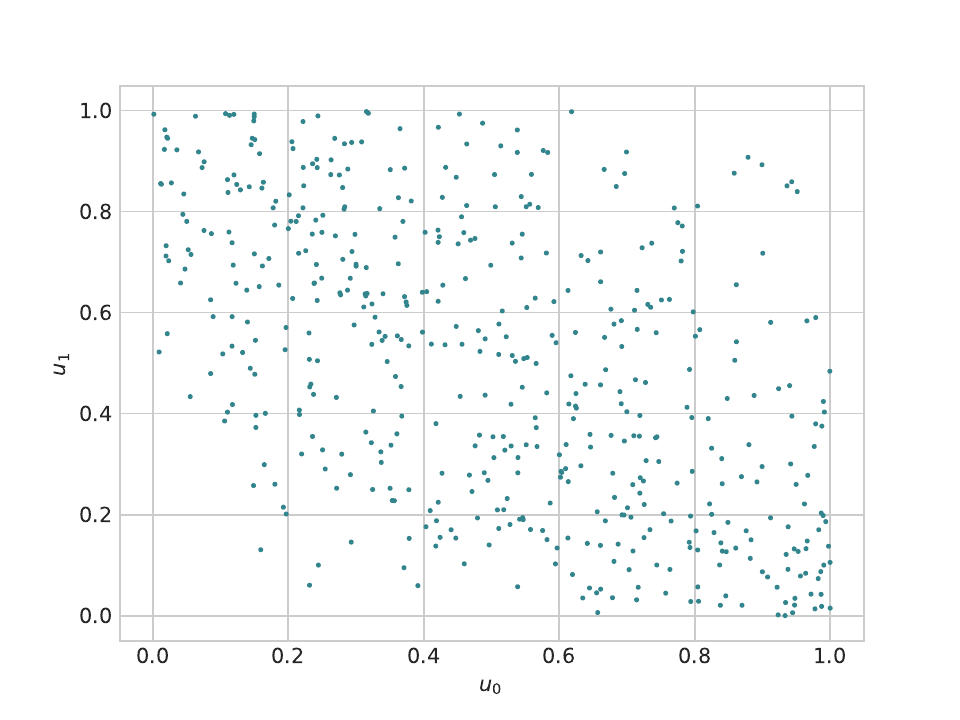"}};
            \draw (1, 1) node {\tiny$\theta = 1$};
        \end{tikzpicture}
        \end{minipage}
      \\
      \textbf{Nelsen n°10} & $\ln(2t^{-\theta}-1)$ & $\theta \in ]0, 1]$ & 
      \begin{minipage}{.3\textwidth}
        \centering
        \begin{tikzpicture}
            \draw (0, 0) node[inner sep=0] {\includegraphics[width=1.5in]{"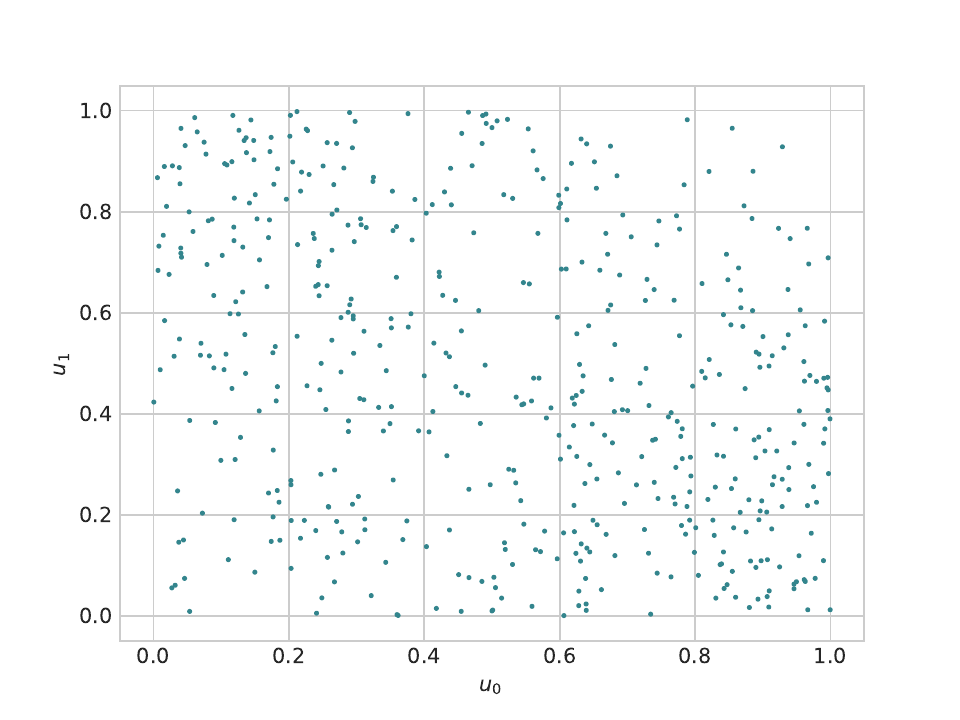"}};
            \draw (1, 1) node {\tiny$\theta = 1$};
        \end{tikzpicture}
        \end{minipage}
      \\
      \hline
  \end{tabular}
\end{table}

\begin{table}[!htp]
	\centering 
	\caption{Bivariate archimedean models in \textbf{COPPY} module.} \label{tab:arch_model_2}
  \begin{tabular}
      {ccccc} \hline Name & $\varphi(t)$ & Constraints & Figure \\
      \hline 
      \textbf{Nelsen n°11} & $\ln(2-t^{\theta})$ & $\theta \in ]0, 0.5]$ &
      \begin{minipage}{.3\textwidth}
        \centering
        \begin{tikzpicture}
            \draw (0, 0) node[inner sep=0] {\includegraphics[width=1.5in]{"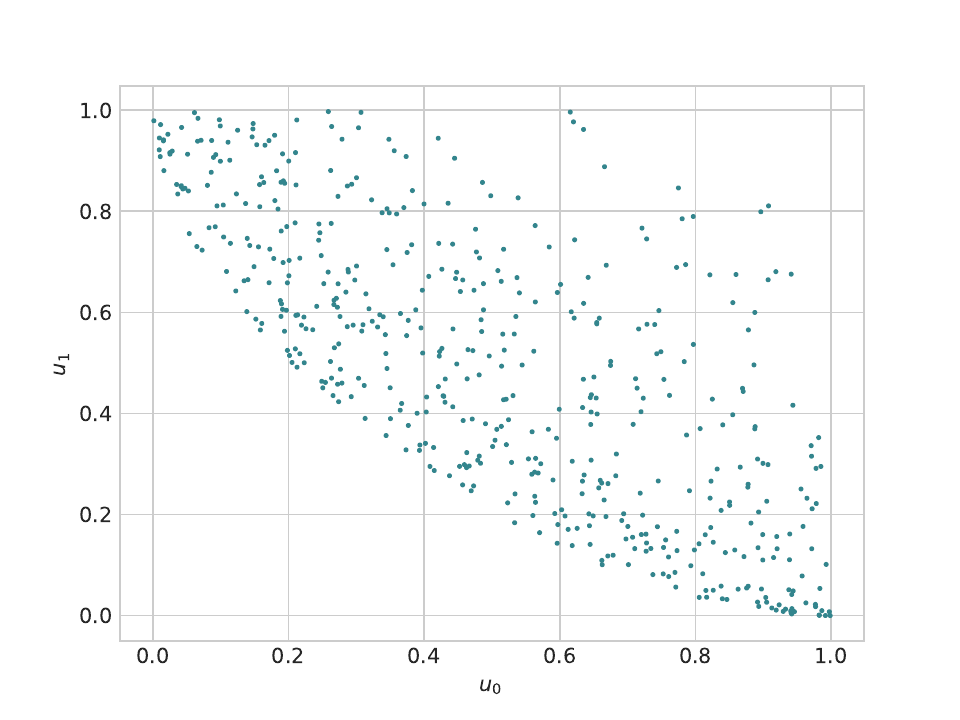"}};
            \draw (1, 1) node {\tiny$\theta = 1.5$};
        \end{tikzpicture}
        \end{minipage}
     \\
      \textbf{Nelsen n°12} & $(\frac{1}{t}-1)^\theta$ & $\theta \in ]0, \infty) \setminus \{0\}$ &
      \begin{minipage}{.3\textwidth}
        \centering
        \begin{tikzpicture}
            \draw (0, 0) node[inner sep=0] {\includegraphics[width=1.5in]{"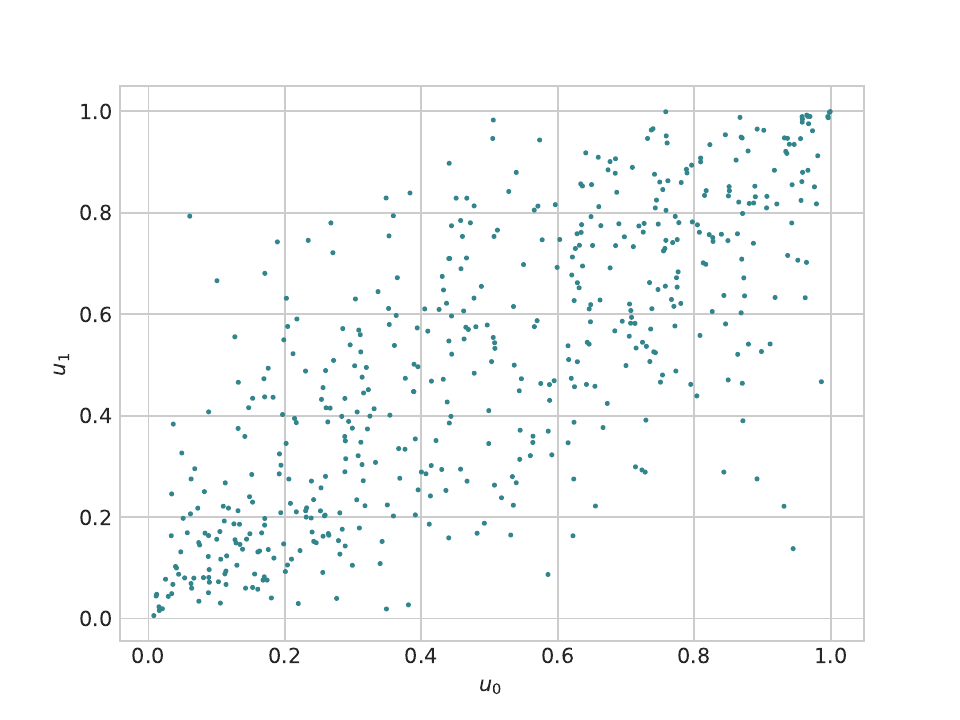"}};
            \draw (1, 1) node {\tiny$\theta = 1.5$};
        \end{tikzpicture}
        \end{minipage}
        \\
      \textbf{Nelsen n°13} & $(1-\ln(t))^\theta-1$ & $\theta \in ]0,\infty[$&
      \begin{minipage}{.3\textwidth}
        \centering
        \begin{tikzpicture}
            \draw (0, 0) node[inner sep=0] {\includegraphics[width=1.5in]{"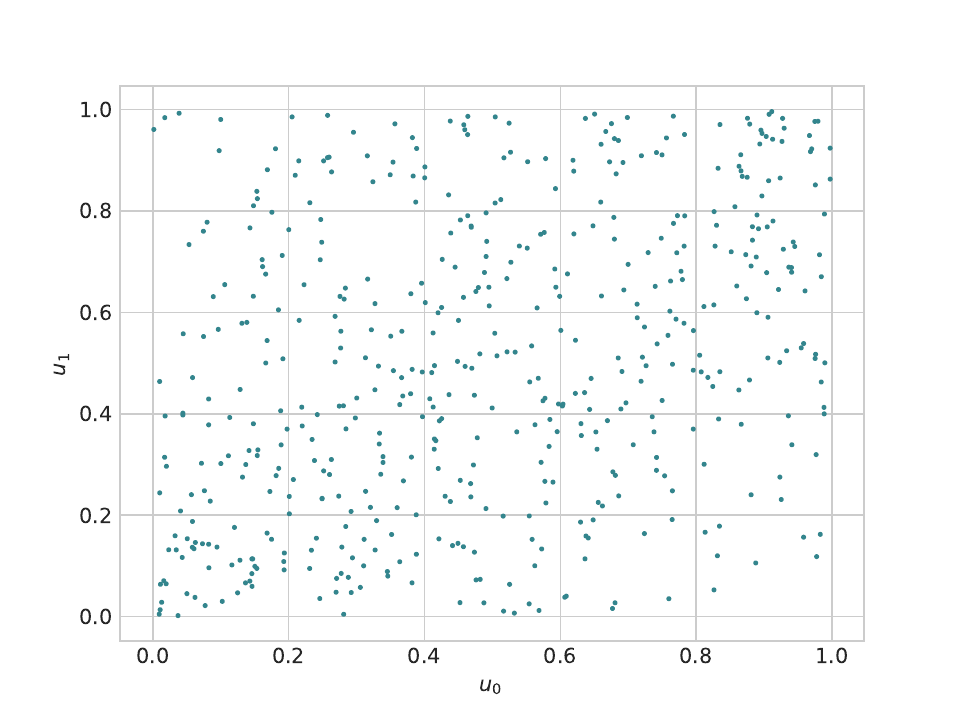"}};
            \draw (1, 1) node {\tiny$\theta = 2$};
        \end{tikzpicture}
        \end{minipage} 
        \\
      \textbf{Nelsen n°14} & $(t^{-\frac{1}{\theta}}-1)^\theta$ & $\theta \in ]1, \infty)$ & 
      \begin{minipage}{.3\textwidth}
        \centering
        \begin{tikzpicture}
            \draw (0, 0) node[inner sep=0] {\includegraphics[width=1.5in]{"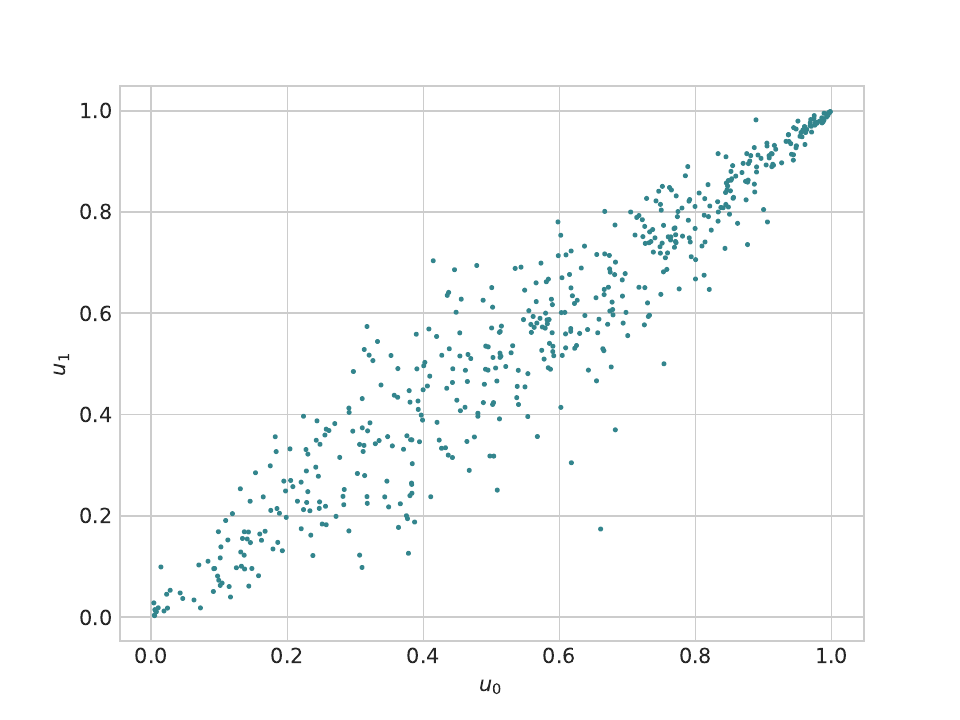"}};
            \draw (1, 1) node {\tiny$\theta = 5$};
        \end{tikzpicture}
        \end{minipage}
      \\
      \textbf{Nelsen n°15} & $(1-t^{\frac{1}{\theta}})^\theta$& $[1, \infty)$ &
      \begin{minipage}{.3\textwidth}
        \centering
        \begin{tikzpicture}
            \draw (0, 0) node[inner sep=0] {\includegraphics[width=1.5in]{"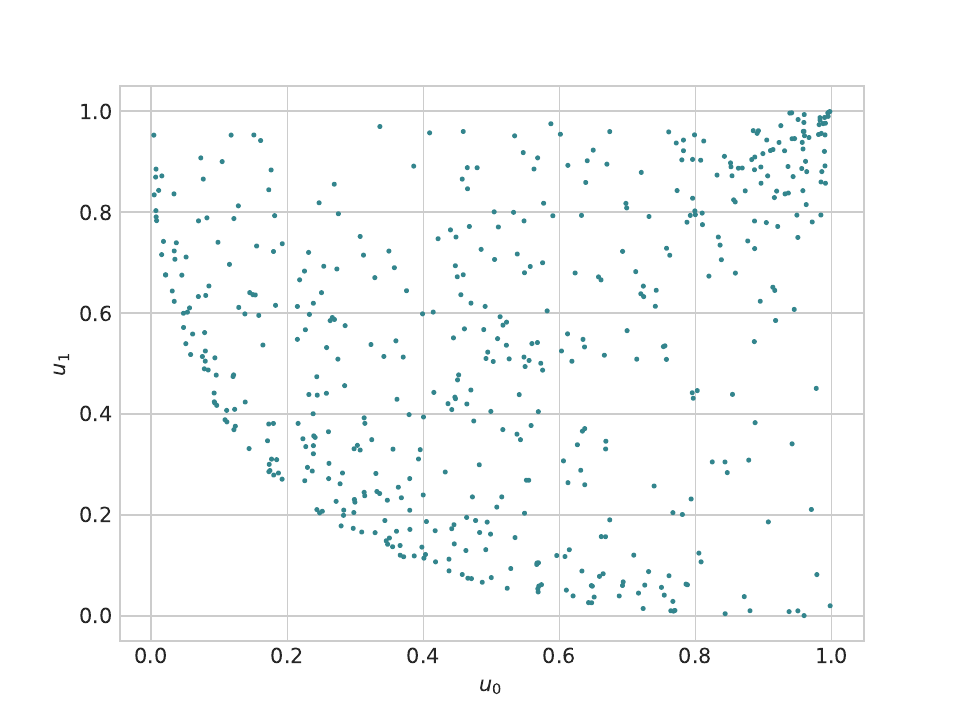"}};
            \draw (1, 1) node {\tiny$\theta = 1.5$};
        \end{tikzpicture}
        \end{minipage}
      \\
      \textbf{Nelsen n°22} & $arcsin(1-t^\theta)$ & $\theta \in [0,1]$ &
      \begin{minipage}{.3\textwidth}
        \centering
        \begin{tikzpicture}
            \draw (0, 0) node[inner sep=0] {\includegraphics[width=1.5in]{"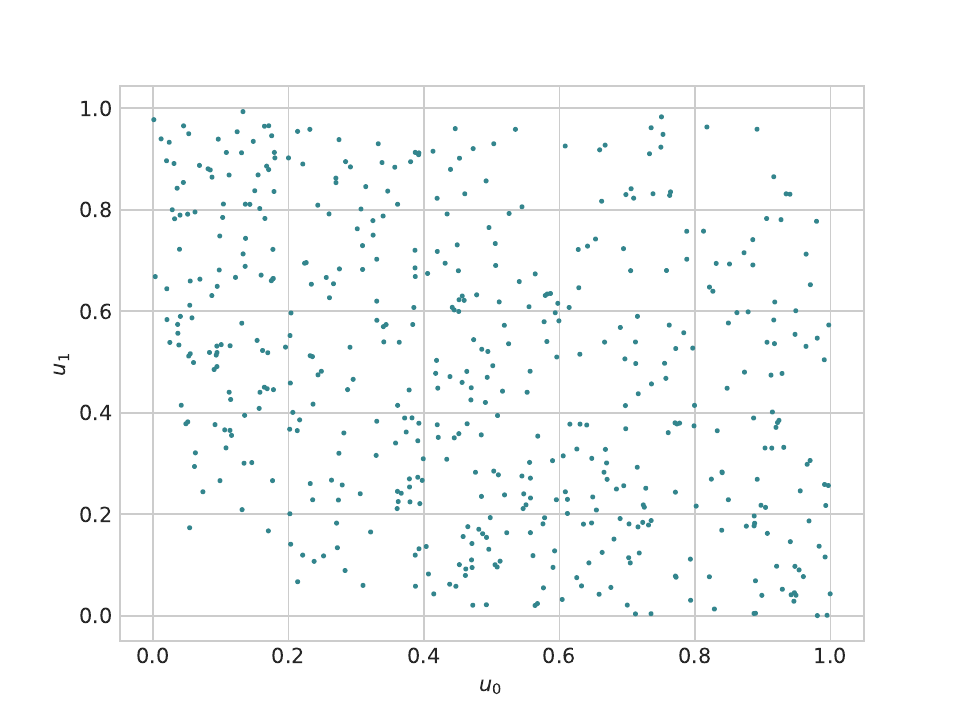"}};
            \draw (1, 1) node {\tiny$\theta = 0.5$};
        \end{tikzpicture}
        \end{minipage}
     \\
      \hline
  \end{tabular}
\end{table}

\clearpage

    \section{Implemented bivariate extreme models} \label{app:bv_ext}

        \begin{table}[!htp]
        \centering
        \caption{Bivariate extreme models in \textbf{COPPY} module.} \label{tab:bv_model}
          \begin{tabular}
              {ccccc} \hline Name & $A(w)$ & Constraints & Figure \\
              \hline \textbf{Gumbel} & $[w^{1/\theta} + (1-w)^{1/\theta}]^{\theta}$ & $\theta \in ]0, 1]$ & 
              \begin{minipage}{.3\textwidth}
                \centering
                \begin{tikzpicture}
                    \draw (0, 0) node[inner sep=0] {\includegraphics[width=1.5in]{"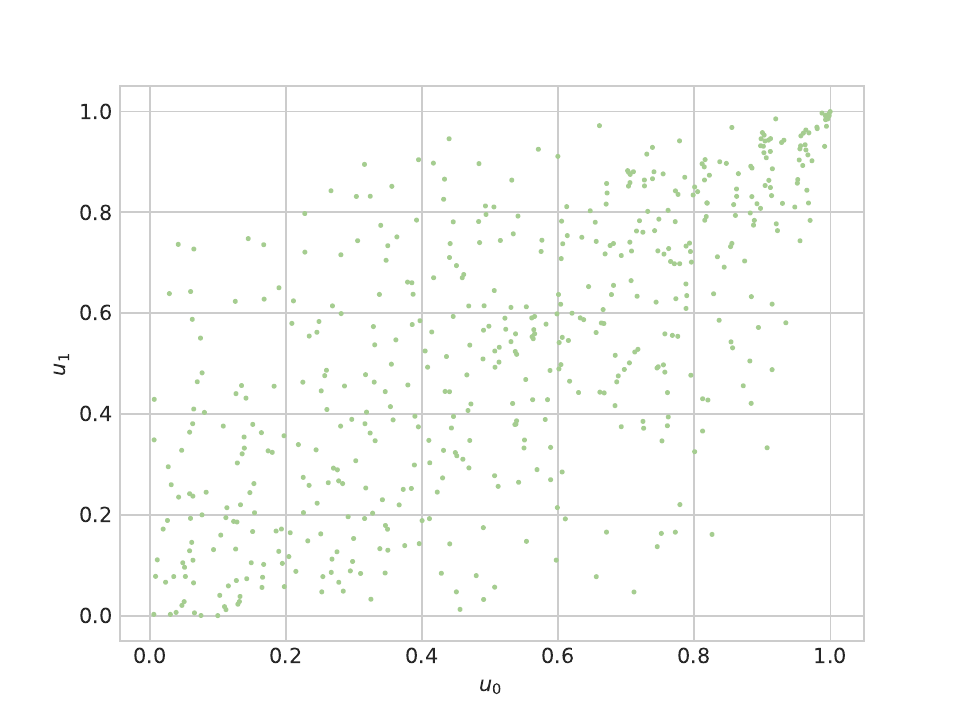"}};
                    \draw (1, 1) node {\tiny$\theta = 0.5$};
                \end{tikzpicture}
              \end{minipage} \\
              \textbf{Galambos} & $1-[w^{-\theta} + (1-w)^{-\theta}]^{-\frac{1}{\theta}}$ & $\theta \in [0, \infty)$ & 
              \begin{minipage}{.3\textwidth}
                \centering
                \begin{tikzpicture}
                    \draw (0, 0) node[inner sep=0] {\includegraphics[width=1.5in]{"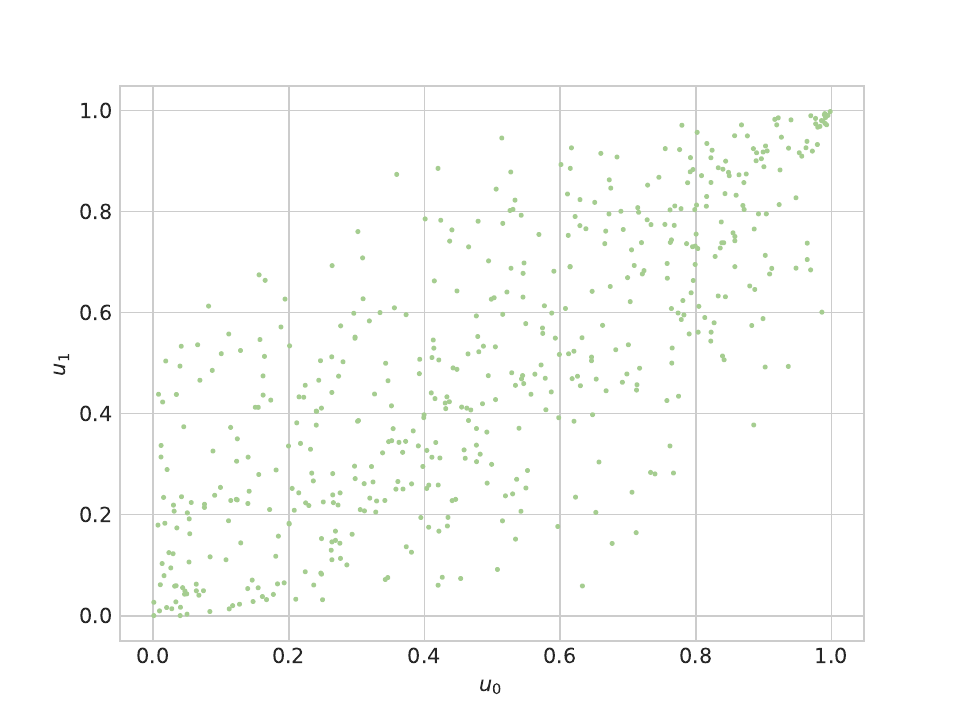"}};
                    \draw (1, 1) node {\tiny$\theta = 1.5$};
                \end{tikzpicture}
              \end{minipage}
                \\
              \textbf{Asy. log.} & \Gape[0pt][12pt]{\makecell{$(1-\psi_1)w + (1-\psi_2)(1-w)$ \\ $+[(\psi_1w)^\theta + (\psi_2(1-w))^\theta]^{\frac{1}{\theta}}$}} & \makecell{$\theta \in [1, \infty)$ \\ $\psi_1, \psi_2 \in (0,1]$} &
              \begin{minipage}{.3\textwidth}
                \centering
                \begin{tikzpicture}
                    \draw (0, 0) node[inner sep=0] {\includegraphics[width=1.5in]{"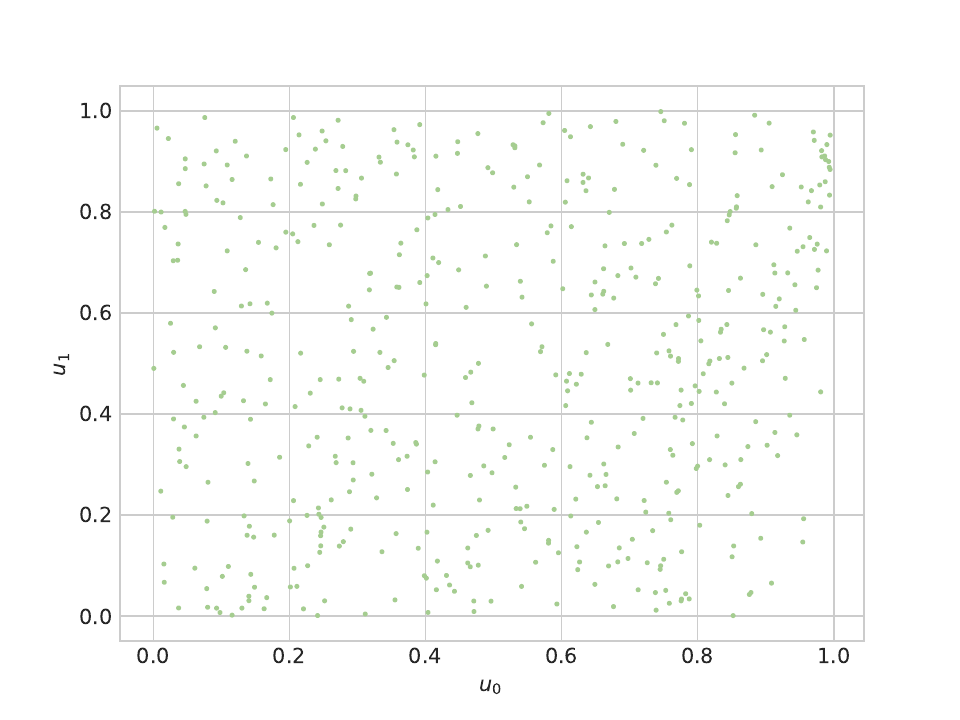"}};
                    \draw (1, 1) node {\tiny$\theta = 1.5$};
                    \draw (1, 0.75) node {\tiny$\psi_1 = 0.1$};
                    \draw (1, 0.5) node {\tiny$\psi_2 = 1.0$};
                \end{tikzpicture}
                \end{minipage}
                \\
              \textbf{\Gape[0pt][12pt]{\makecell{Asy. neg.\\ log.}}} & $1-[(\psi_1w)^{-\theta} + (\psi_2(1-w))^{-\theta}]^{-\frac{1}{\theta}}$ & \makecell{$\theta \in [0, \infty)$ \\ $\psi_1, \psi_2 \in (0,1]$} & 
              \begin{minipage}{.3\textwidth}
                \centering
                \begin{tikzpicture}
                    \draw (0, 0) node[inner sep=0] {\includegraphics[width=1.5in]{"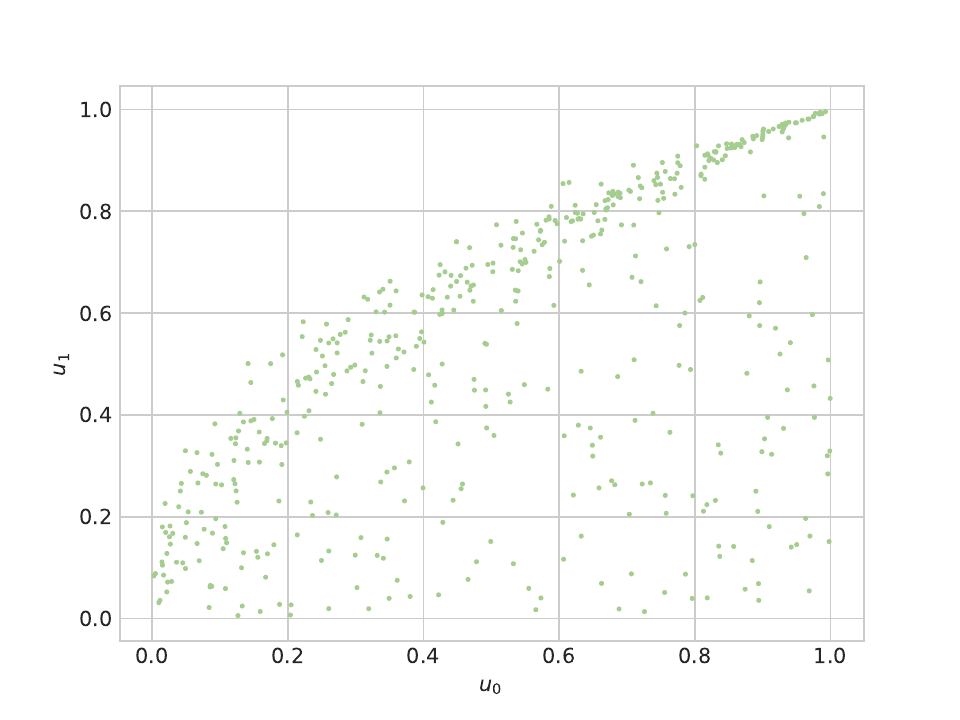"}};
                    \draw (-1, 1) node {\tiny $\theta = 10$};
                    \draw (-1, 0.75) node {\tiny $\psi_1 = 0.5$};
                    \draw (-1, 0.5) node {\tiny $\psi_2 = 1.0$};
                \end{tikzpicture}
                
              \end{minipage}  \\
              \textbf{Asy. mixed} & $1-(\theta+\psi_1)w+\theta w^2 + \psi_1 w^3$ & \Gape[0pt][12pt]{\makecell{$\theta \geq 0$, \\ $\theta + 3\psi_1 \geq 0$, \\ $\theta + \psi_1 \leq 1$, \\ $\theta+2\psi_1 \leq 1$.}} &
              \begin{minipage}{.3\textwidth}
                \centering
                \begin{tikzpicture}
                    \draw (0, 0) node[inner sep=0] {\includegraphics[width=1.5in]{"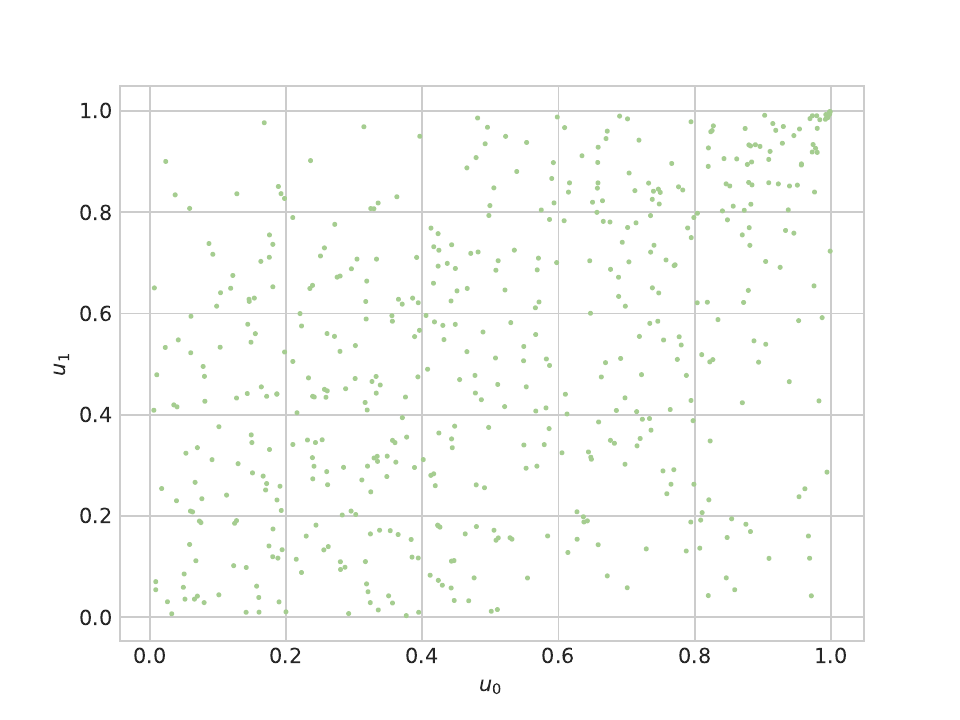"}};
                    \draw (1, 1) node {\tiny$\theta = 4/3$};
                    \draw (1, 0.75) node {\tiny$\psi_1 = -1/3$};
                \end{tikzpicture}
              \end{minipage}
                \\
              \textbf{Husler Reiss} & \Gape[0pt][12pt]{\makecell{$(1-w)\Phi(\theta +\frac{1}{2\theta}\ln(\frac{1-w}{w}))$ \\ $+w\Phi(\theta +\frac{1}{2\theta}\ln(\frac{w}{1-w}))$}} & $\theta \in (0, \infty)$ & \begin{minipage}{.3\textwidth}
                \centering
                \begin{tikzpicture}
                    \draw (0, 0) node[inner sep=0] {\includegraphics[width=1.5in]{"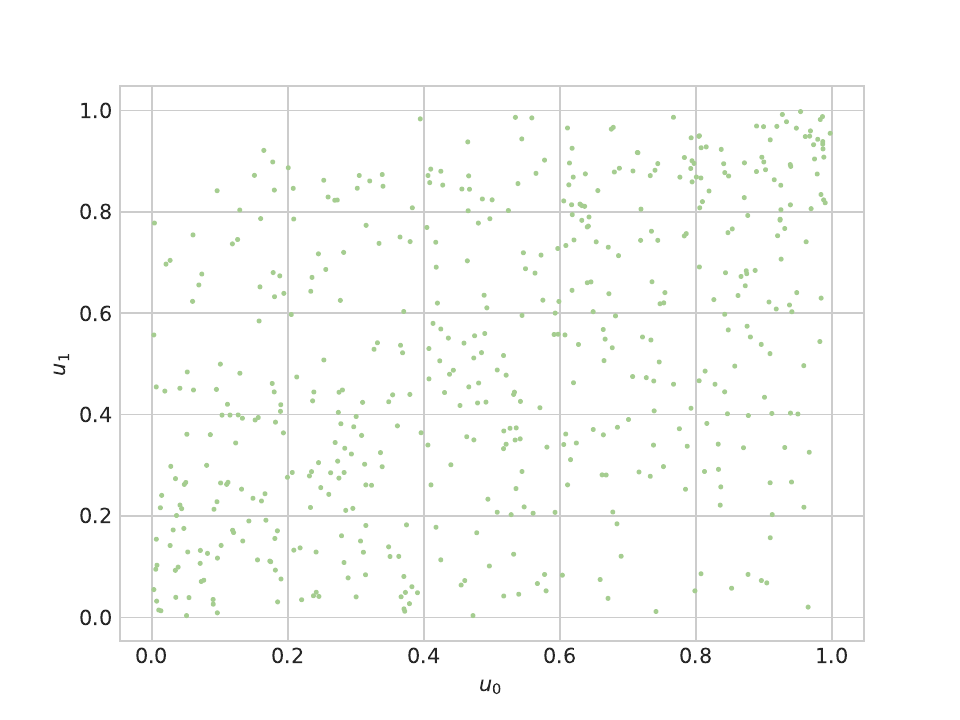"}};
                    \draw (1, 1) node {\tiny$\theta = 1.0$};
                \end{tikzpicture}
              \end{minipage}
               \\
              \textbf{$t$-EV} & $wt_{\psi_1+1}(z_{w}) + (1-w)t_{\psi_1+1}(z_{1-w})$ & $\theta \in (-1, 1)$ & 
              \begin{minipage}{.3\textwidth}
                \centering
                \begin{tikzpicture}
                    \draw (0, 0) node[inner sep=0] { \includegraphics[width=1.5in]{"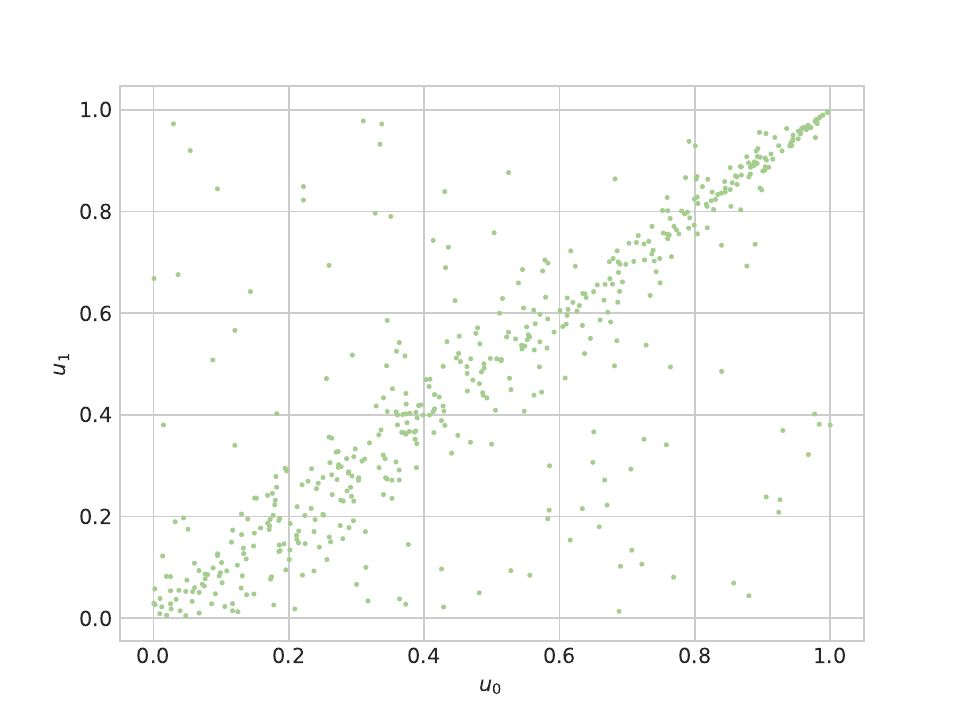"}};
                    \draw (1, 1) node {\tiny$\theta = 0.8$};
                    \draw (1, 0.75) node {\tiny$\psi_1 = 0.2$};
                \end{tikzpicture}
              \end{minipage}
                \\
              \hline
          \end{tabular}
        \end{table}

\clearpage

\section{Multivariate Archimedean copulae}
    \label{app:mv_arch}
	\begin{table}[!htp]
        \centering
        \caption{Multivariate archimedean models in \textbf{COPPY} module.} \label{tab:mv_arch_model_1}
          \begin{tabular}
              {ccccc} \hline Name & $\varphi(t)$ & Constraints & Figure \\
              \hline \textbf{Clayton} & $\frac{1}{\theta}(t^{-\theta}-1)$ & $\theta \in [0, \infty) \setminus \{0\}$ &
              \begin{minipage}{.3\textwidth}
                \centering
                \begin{tikzpicture}
                    \draw (0, 0) node[inner sep=0] {\includegraphics[width=1.75in]{"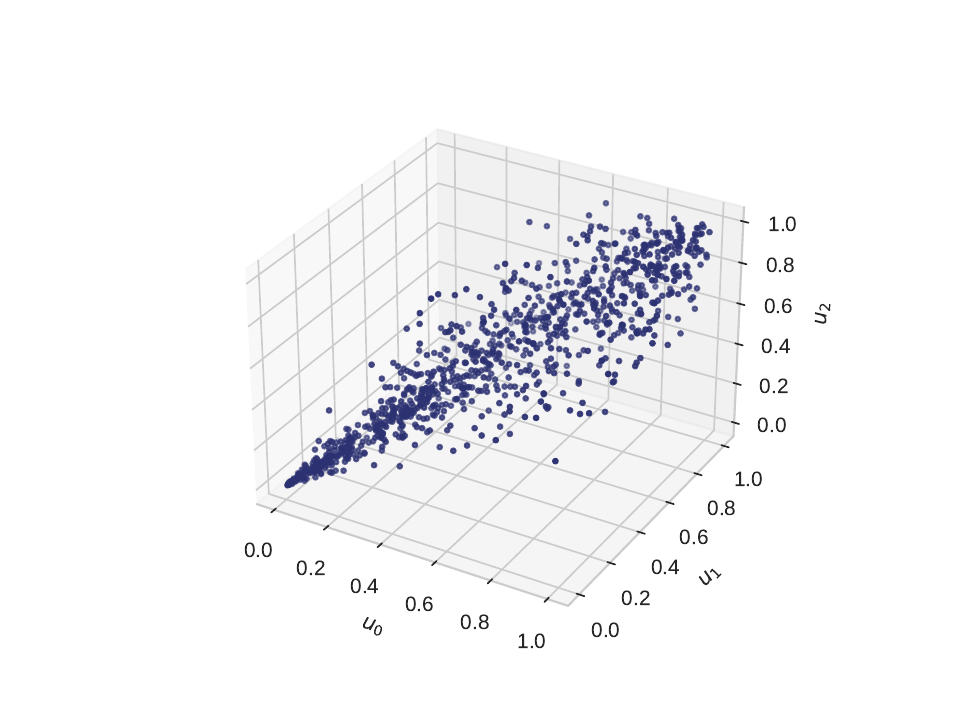"}};
                    \draw (1, 1) node {\tiny$\theta = 5$};
                \end{tikzpicture}
                \end{minipage}
               \\
              \textbf{AMH} & $\ln(\frac{1-\theta(1-t)}{t})$ & $\theta \in [-1, 1)$ &
              \begin{minipage}{.3\textwidth}
                \centering
                \begin{tikzpicture}
                    \draw (0, 0) node[inner sep=0] {\includegraphics[width=1.75in]{"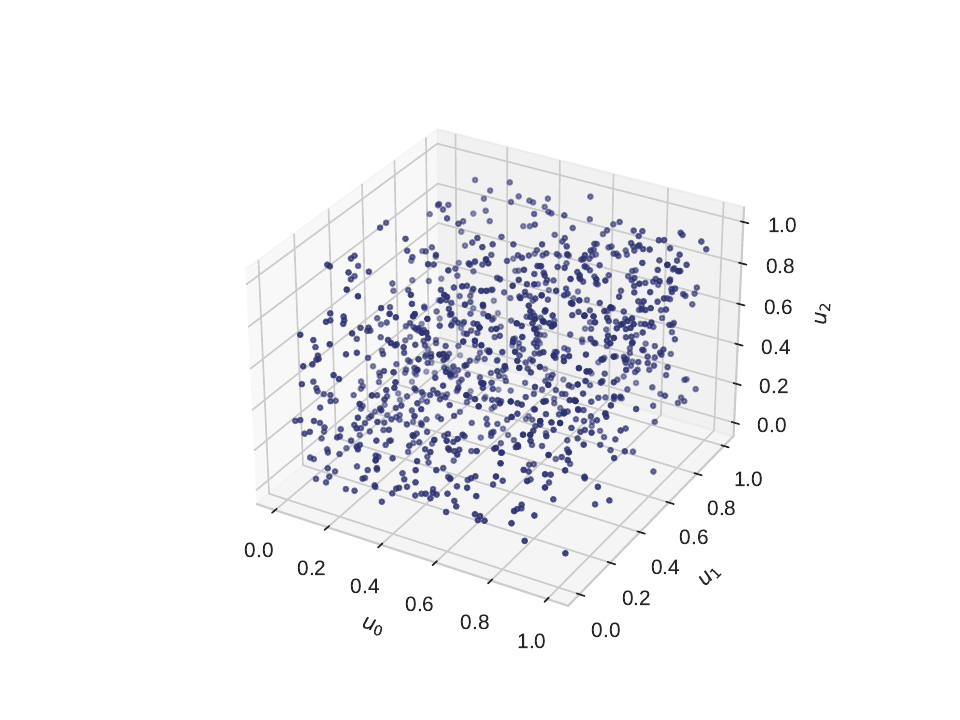"}};
                    \draw (1, 1) node {\tiny$\theta = 0.5$};
                \end{tikzpicture}
                \end{minipage}  \\
              \textbf{Frank} & $-\ln(\frac{e^{-\theta t}-1}{e^{-\theta}-1})$ & $\theta \in \mathbb{R}\setminus \{0\}$ &
              \begin{minipage}{.3\textwidth}
                \centering
                \begin{tikzpicture}
                    \draw (0, 0) node[inner sep=0] {\includegraphics[width=1.75in]{"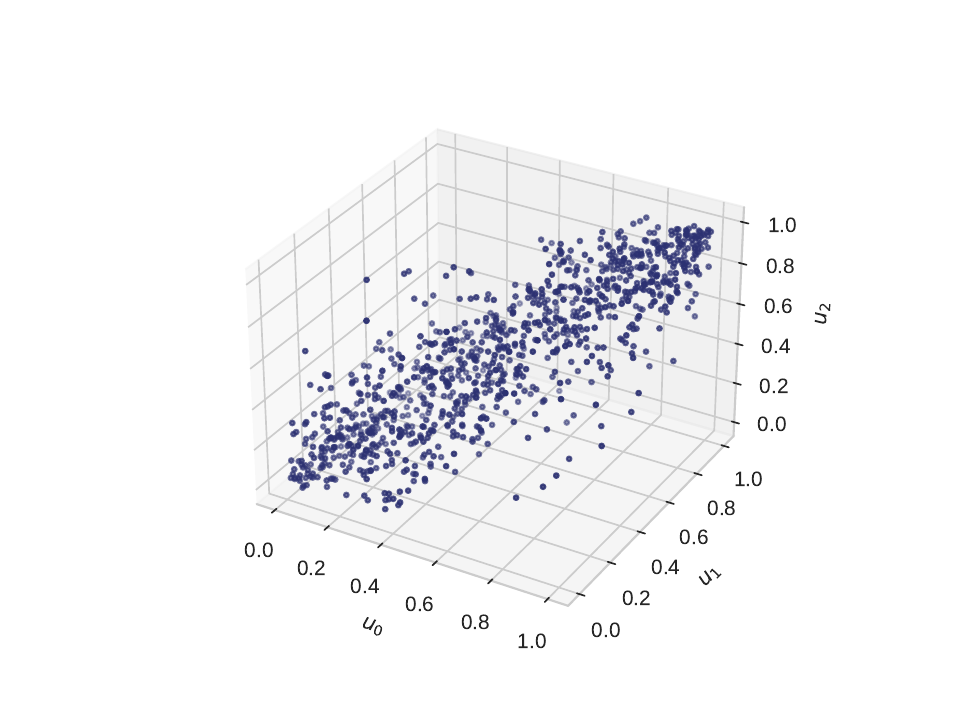"}};
                    \draw (1, 1) node {\tiny$\theta = 8$};
                \end{tikzpicture}
                \end{minipage}
              \\
              \textbf{Joe} & $-\ln(1-(1-t)^\theta)$ & $\theta \in [1, \infty)$ &
              \begin{minipage}{.3\textwidth}
                \centering
                \begin{tikzpicture}
                    \draw (0, 0) node[inner sep=0] {\includegraphics[width=1.75in]{"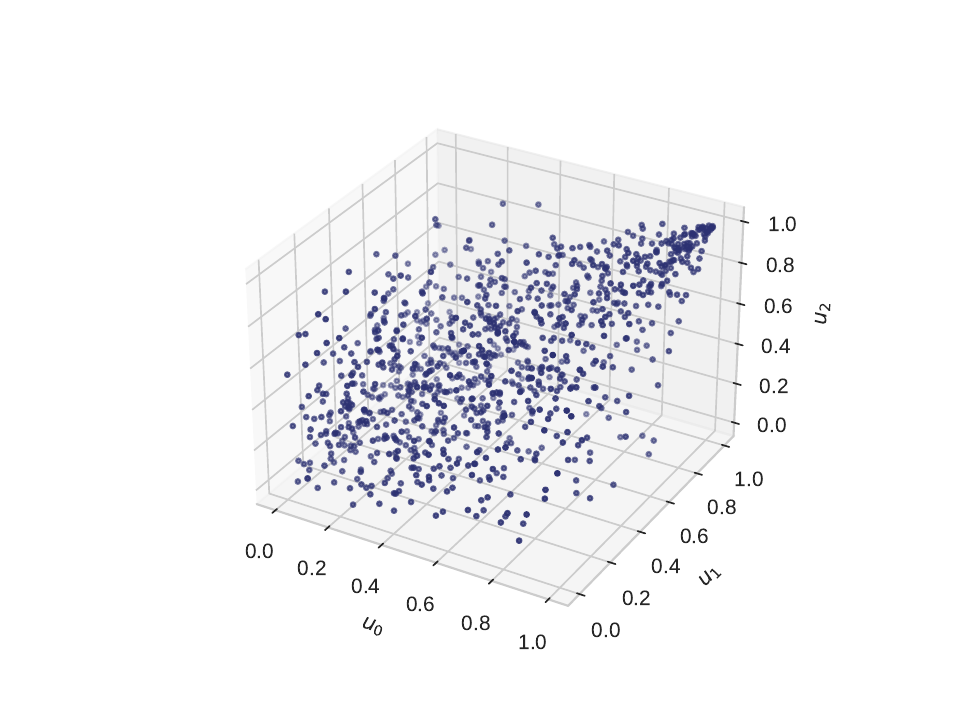"}};
                    \draw (1, 1) node {\tiny$\theta = 2$};
                \end{tikzpicture}
                \end{minipage}
                \\
              \hline
          \end{tabular}
        \end{table}
	
\clearpage

    \section{Multivariate extreme models}
        \label{app:mv_ext}
        Before giving the main details, we introduce some notations. Let $B$ be the set of all nonempty subsets of $\{1,\dots,d\}$ and $B_1 = \{b \in B, |b| = 1\}$, where $|b|$ denotes the number of elements in thet set $b$. We note by $B_{(j)} = \{b \in B, j \in b\}$. For $d=3$, the Pickands is expressed as 
        \begin{align*}
            A(\textbf{w}) =& \alpha_1 w_1 + \psi_1 w_2 + \phi_1 w_3 + \left( (\alpha_2 w_1)^{\theta_1} + (\psi_2w_2)^{\theta_1} \right)^{1/\theta_1} + \left( (\alpha_3 w_2)^{\theta_2} + (\phi_2w_3)^{\theta_2} \right)^{1/\theta_2} \\ &+ \left( (\psi_3 w_2)^{\theta_3} + (\phi_3w_3)^{\theta_3} \right)^{1/\theta_3} 
                                  + \left( (\alpha_4 w_1)^{\theta_4} + (\psi_4 w_2)^{\theta_4} + (\phi_4 w_3)^{\theta_4} \right)^{1/\theta_4},
            \end{align*}
        where $\boldsymbol{\alpha} = (\alpha_1, \dots, \alpha_4), \boldsymbol{\psi} = (\psi_1, \dots, \psi_4), \boldsymbol{\phi} = (\phi_1, \dots, \phi_4)$ are all elements of $\Delta^3$. We take $\boldsymbol{\alpha} = (0.4,0.3,0.1,0.2)$, $\boldsymbol{\psi} = (0.1, 0.2, 0.4, 0.3)$, $\boldsymbol{\phi} = (0.6,0.1,0.1,0.2)$ and $\boldsymbol{\theta} = (\theta_1, \dots, \theta_4) = (0.6,0.5,0.8,0.3)$ as the dependence parameter.
        
        The Dirichlet model is a mixture of $m$ Dirichlet densities, that is
        \begin{equation*}
            h(\textbf{w}) = \sum_{k=1}^m \theta_k \frac{\Gamma(\sum_{j=1}^d \sigma_{kj})}{\Pi_{j=1}^d \Gamma(\sigma_{kj})} \Pi_{j=1}^d w_j^{\sigma_{kj}-1},
        \end{equation*}
        with $\sum_{k=1}^m \theta_k = 1$, $\sigma_{kj} > 0$ for $k \in \{1,\dots,m\}$ and $j \in \{1, \dots, d\}$. Let $\mathcal{D} \in [0, \infty)^{(d-1)\times (d-1)}$ denotes the space of symmetric strictly conditionnaly negative definite matrices that is
        \begin{align*}
            \mathcal{D}_{k} = \Big\{ \Gamma \in [0,\infty)^{k \times k} : a^\top \Gamma a < 0 \; \textrm{for all } a \in \mathbb{R}^{k} \setminus \{\textbf{0}\}  \, \textrm{with } \sum_{j=1}^{d-1} a_j = 0, \\ \Gamma_{ii} = 0, \Gamma_{ij} = \Gamma_{ji}, \quad 1 \leq i,j\leq k \Big\}.
        \end{align*}
        For any $2 \leq k \leq d$ consider $m' = (m_1, \dots, m_k)$ with $1 \leq m_1 <  \dots < m_k \leq d$ define
        \begin{equation*}
            \Sigma^{(k)}_m = 2 \left( \Gamma_{m_i m_k} + \Gamma_{m_j m_k} - \Gamma_{m_i m_j} \right)_{m_i m_j \neq m_k} \in [0,\infty)^{(d-1)\times(d-1)}.
        \end{equation*}
        Furthermore, note $S(\cdot | \Sigma^{(k)}_m)$ denote the survival function of a normal random vector with mean vector $\textbf{0}$ and covariance matrix $\Sigma^{(k)}$. We now define :
        \begin{equation*}
            h_{km}(\textbf{y}) = \int_{y_k}^\infty S\left( (y_i - z + 2\Gamma_{m_i m_k})_{i=1}^{k-1} | \Gamma_{km}\right) e^{-z}dz 
        \end{equation*}
        for $2 \leq k \leq d$. We denote by $\Sigma^{(k)}$ the summation over all $k$-vectors $m=(m_1,\dots,m_k)$ with $1\leq m_1 < \dots < m_k \leq d$.
    
        \begin{table}[!htp]
        	\centering 
        	\caption{Multivariate extreme models in \textbf{COPPY} module.} \label{tab:mv_ext_mod}
          \begin{tabular}
              {ccccc} \hline Name & $A(\textbf{w})$ & Constraints & Figure \\
              \hline \textbf{Logistic} & $\left(\sum_{j=1}^d w_j^\frac{1}{\theta}\right)^{\theta}$ & $\theta \in ]0, 1]$ &
              \begin{minipage}{.3\textwidth}
                \centering
                \begin{tikzpicture}
                    \draw (0, 0) node[inner sep=0] {\includegraphics[width=1.75in]{"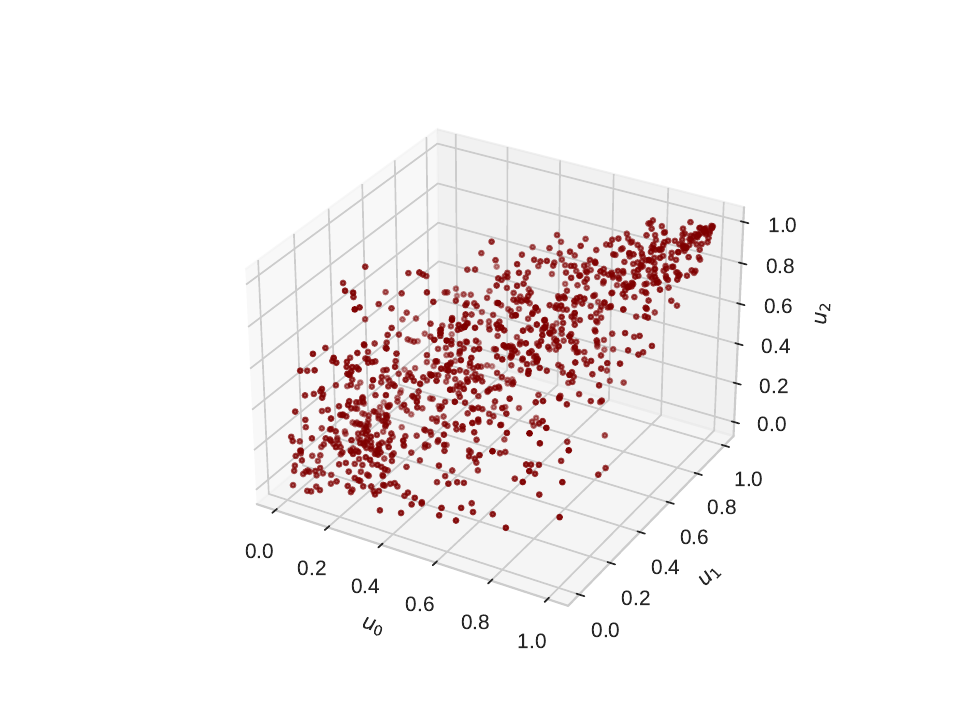"}};
                    \draw (1, 1) node {\tiny$\theta = 0.5$};
                \end{tikzpicture}
                \end{minipage} 
                \\
              \textbf{Asy. Log.} & $\sum_{b \in B}( \sum_{j \in b} (\psi_{j,b}w_j)^{\frac{1}{\theta_b}})^{\theta_b}$ & \makecell{$\theta_b \in ]0, 1] \, \forall b \in B \setminus B_1$,\\ $\psi_{j,b} \in [0,1] \, \forall b \in B \, \forall j \in b$, \\ $\sum_{b \in B_{(j)}} \psi_{j,b} = 1$, $j \in [\![d-1]\!]$, \\ $\theta_b = 1 \, \forall b \in B \setminus B_1 \implies$ \\ $\psi_{j,b} = 0 \, \forall j \in b$.} &
              \begin{minipage}{.3\textwidth}
                \centering
                \begin{tikzpicture}
                    \draw (0, 0) node[inner sep=0] {\includegraphics[width=1.75in]{"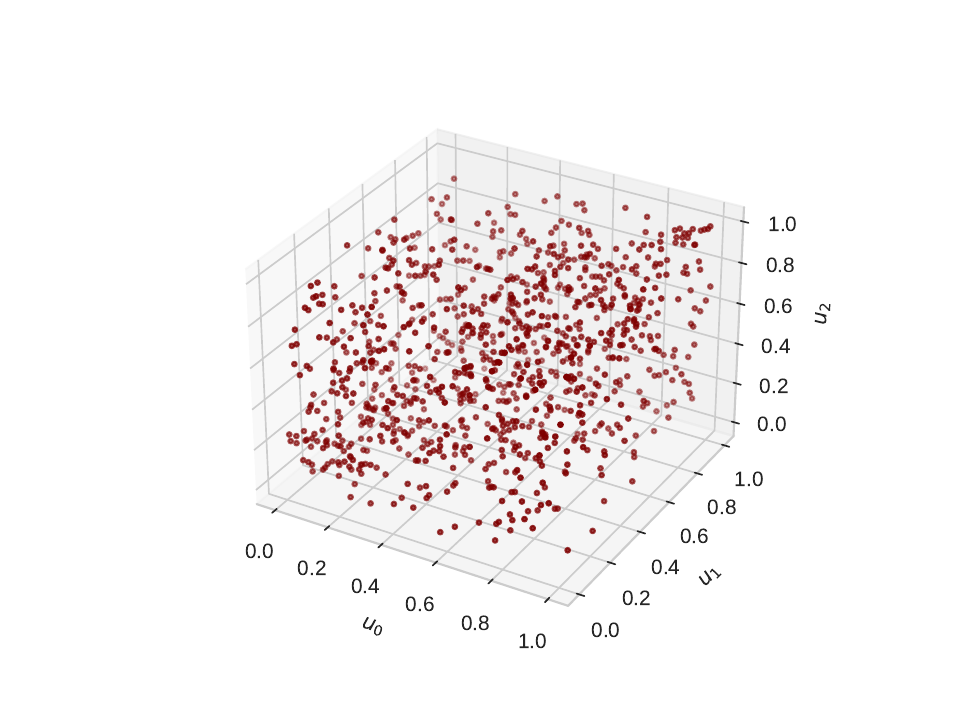"}};
                \end{tikzpicture}
                \end{minipage}  
                \\
                \textbf{Dirichlet} & Not specified & \Gape[0pt][12pt]{\makecell{$\sum_{k=1}^m \theta_k = 1$, \\ $\sigma_{kj} > 0, k \in \{1,\dots,m\}$, \\ $j \in \{1,\dots,d\}$}} &
              \begin{minipage}{.3\textwidth}
                \centering
                \begin{tikzpicture}
                    \draw (0, 0) node[inner sep=0] {\includegraphics[width=1.75in]{"figures/mv_plot/mv_asy_log.pdf"}};
                    \draw (1, 1) node {\tiny$\theta = (1/3,1/3,1/3)$};
                    \draw (1, -1) node {\tiny $ \Sigma = \begin{pmatrix} 2 & 1 & 1\\ 1 & 2 & 1 \\ 1 & 1 & 2  \end{pmatrix}$};
                \end{tikzpicture}
                \end{minipage}  
                \\
                \textbf{Hüsler Reiss} & \Gape[0pt][12pt]{\makecell{$\sum_{k=1}^d (-1)^{k+1} \times$ \\ $\Sigma^{(k)} h_{km} (u_{m_1}, \dots, u_{m_k})$}} & $\Gamma \in \mathcal{D}_d$ &
              \begin{minipage}{.3\textwidth}
                \centering
                \begin{tikzpicture}
                    \draw (0, 0) node[inner sep=0] {\includegraphics[width=1.75in]{"figures/mv_plot/mv_asy_log.pdf"}};
                    \draw (1, -1) node {\tiny $ \Gamma = \begin{pmatrix} 0 & 3 & 3\\ 3 & 0 & 3 \\ 3 & 3 & 0  \end{pmatrix}$};
                \end{tikzpicture}
                \end{minipage} 
                \\
              \hline
          \end{tabular}
        \end{table}

\section{Multivariate elliptical dependencies} \label{app:mv_ellip}

Let $\textbf{X} \sim \textbf{E}_d(\boldsymbol{\mu}, \Sigma, \psi)$ be an elliptical distributed random vector with cumulative distribution $F$ and marginal $F_0, \dots, F_{d-1}$. Then, the copula $C$ of $F$ is called an elliptical copula. We denote by  $\phi$ the standard normal distribution function and $\boldsymbol{\phi}_\Sigma$ the joint distribution function of $\textbf{X} \sim \mathcal{N}_d(\textbf{0}, \Sigma)$, where $\textbf{0}$ is the $d$-dimensional vector composed out of $0$. In the same way, we note $t_{\theta}$ the distribution function of a standard univariate distribution $t$ distribution and by $\boldsymbol{t}_{\theta, \Sigma}$ the joint distribution function of the vector $\textbf{X} \sim \boldsymbol{t}_{d}(\theta, \textbf{0}, \Sigma)$. A $d$ squared matrix $\Sigma$ is said to be positively semi definite if for all $u \in \mathbb{R}^d$ we have :

\begin{equation*}
    u^\top \Sigma u \geq 0
\end{equation*}

\begin{table}[!htp]
	\centering 
	\caption{Multivariate elliptical models in \textbf{COPPY} module.} \label{tab:mv_ell_mod}
  \begin{tabular}
      {ccccc} \hline Name & C & Constraints & Figure \\
      \hline \textbf{Gaussian} & $\boldsymbol{\phi}_{\Sigma}(\phi^\leftarrow(u_0),\dots, \phi^\leftarrow(u_{d-1}))$ & $\Sigma$ PSD &
      \begin{minipage}{.3\textwidth}
        \centering
        \begin{tikzpicture}
            \draw (0, 0) node[inner sep=0] {\includegraphics[width=1.75in]{"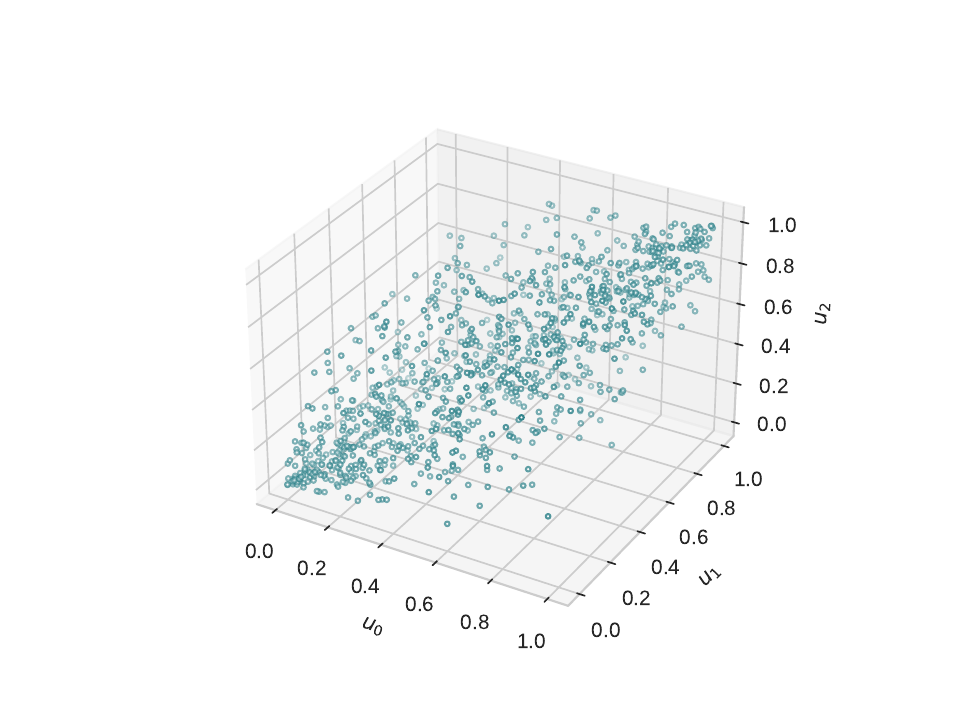"}};
            \draw (1, 1) node {\tiny$\rho = 0.71$};
        \end{tikzpicture}
        \end{minipage}
      \\
      \textbf{Student} & $\boldsymbol{t}_{\theta,\Sigma}(t_{\theta}^\leftarrow(u_0),\dots,t_{\theta}^\leftarrow(u_{d-1}))$ & $\theta > 0$, $\Sigma$ PSD &
      \begin{minipage}{.3\textwidth}
        \centering
        \begin{tikzpicture}
            \draw (0, 0) node[inner sep=0] {\includegraphics[width=1.75in]{"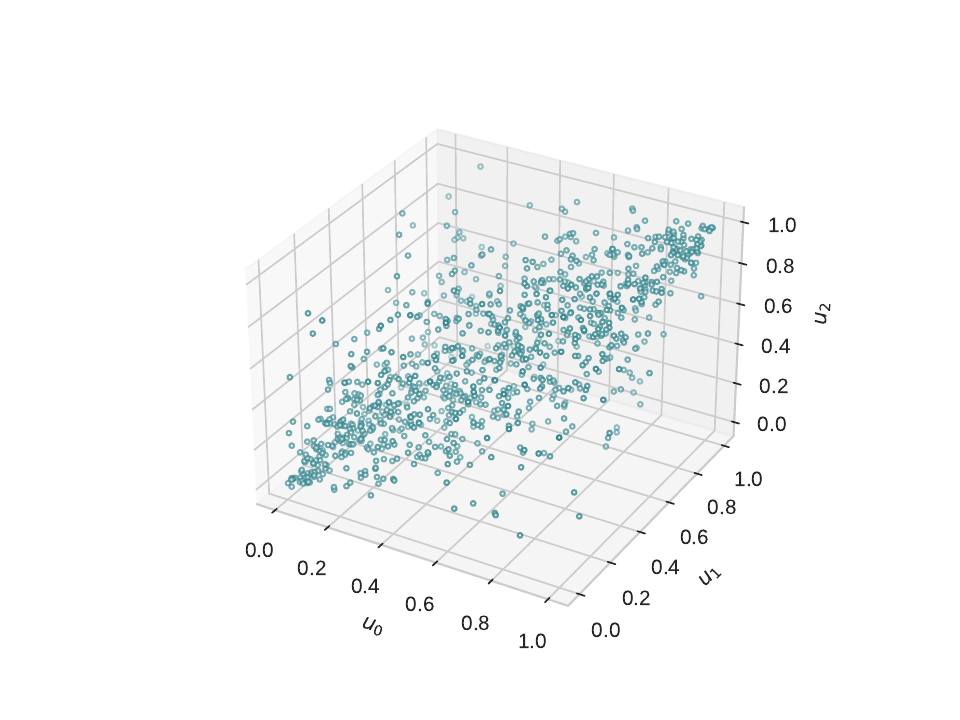"}};
            \draw (1, 1) node {\tiny$\rho = 0.71$};
            \draw (1, 0.75) node {\tiny $\theta = 4$};
        \end{tikzpicture}
        \end{minipage} 
        \\
      \hline
  \end{tabular}
\end{table}

\end{document}